\RequirePackage{fix-cm}
\documentclass[smallcondensed]{svjour3}
\usepackage{graphicx}
\usepackage{caption}
\usepackage{subcaption}

\usepackage{amsmath}
\usepackage{amssymb}

\usepackage{amsthm}

\usepackage{booktabs}

\usepackage{enumitem}

\usepackage{isabelle,isabellesym}
\usepackage{mathtools}
\usepackage{url}

\usepackage{algorithm}
\usepackage{algpseudocode}

\usepackage{booktabs}

\usepackage{tikz}
\tikzset{every picture/.style={remember picture}}
\usetikzlibrary{decorations.markings}
\tikzset{
	set arrow inside/.code={\pgfqkeys{/tikz/arrow inside}{#1}},
	set arrow inside={end/.initial=>, opt/.initial=},
	/pgf/decoration/Mark/.style={
		mark/.expanded=at position #1 with
		{
			\noexpand\arrow[\pgfkeysvalueof{/tikz/arrow inside/opt}]{\pgfkeysvalueof{/tikz/arrow inside/end}}
		}
	},
	arrow inside/.style 2 args={
		set arrow inside={#1},
		postaction={
			decorate,decoration={
				markings,Mark/.list={#2}
			}
		}
	},
}

\usepackage{hyperref}

\newcommand{\blank}{{-}}

\renewcommand{\Re}{\operatorname{Re}}
\renewcommand{\Im}{\operatorname{Im}}
\newcommand{\Ln}{\operatorname{Ln}}
\newcommand{\Ind}{\operatorname{Ind}}
\newcommand{\Indp}{\operatorname{Indp}}

\newcommand{\jump}{\mathrm{jump}}

\newcommand{\TaQ}{\mathrm{TaQ}}
\newcommand{\SRemS}{\mathrm{SRemS}}

\newcommand{\Var}{\mathrm{Var}}

\renewcommand{\div}{\mathbin{\mathrm{div}}}

\newcommand{\sgn}{\mathrm{sgn}}

\newcommand{\Res}[0]{\textrm{Res}}

\begin{document}

\titlerunning{Evaluating Winding Numbers and Counting Complex Roots}
\title{Evaluating Winding Numbers and Counting Complex Roots through Cauchy Indices in Isabelle/HOL
\thanks{
        The first author was funded by the China Scholarship Council, via the CSC Cambridge Scholarship programme.
        This development is also supported by the European Research Council Advanced Grant ALEXANDRIA (Project 742178).}}


\author{Wenda Li  \and Lawrence C. Paulson}

\institute{Wenda Li \at
        Computer Laboratory, University of Cambridge \\
        \email{wl302@cam.ac.uk}           
        \and
        Lawrence C. Paulson \at
        Computer Laboratory, University of Cambridge \\
        \email{lp15@cam.ac.uk}           
}


\maketitle

\begin{abstract}
In complex analysis, the winding number measures the number of times a path (counter-clockwise) winds around a point, while the Cauchy index can approximate how the path winds. We formalise this approximation in the Isabelle theorem prover, and provide a tactic to evaluate winding numbers through Cauchy indices. By further combining this approximation with the argument principle, we are able to make use of remainder sequences to effectively count the number of complex roots of a polynomial within some domains, such as a rectangular box and a half-plane.

\end{abstract}

\section{Introduction}

The winding number, given by
\[
n(\gamma,z) = \frac{1}{2 \pi i} \oint_\gamma \frac{d w}{w - z},
\]
measures how the path $\gamma$ winds around the complex point $z$. It is an important object in complex analysis, and its evaluation is ubiquitous among analytic proofs.

However, when formally evaluating the winding number in proof assistants such as Isabelle/HOL and HOL Light, unexpected difficulties arise, as pointed out by Harrison \cite{harrison-pnt} and Li et al. \cite{Li_ITP2016}.
To address this problem, we formalise a theory of the Cauchy index on the complex plane, thereby approximating how the path winds. When the path is a cycle and comprises line segments and parts of circles, we can now evaluate the winding number by calculating Cauchy indices along those sub-paths.

In addition, by further combining our previous formalisation of the argument principle \cite{Li_ITP2016} (which associates the winding number with the number of complex roots), we build effective procedures to count the complex roots of a polynomial within some domains, such as a rectangle box and a half-plane.

In short, the main contributions of this paper are
\begin{itemize}
	\item a novel tactic to enable users to evaluate the winding number through Cauchy indices,
	\item and novel verified procedures to count complex roots of a polynomial.
\end{itemize}
The Isabelle sources of this paper are available from the Archive of Formal Proofs \cite{Count_Complex_Roots-AFP,Winding_Number_Eval-AFP}.

Formulations in this paper, such as the definition of the Cauchy index and statements of some key lemmas, mainly follow Rahman and Schmeisser's book \cite[Chapter 11]{Rahman:2016us} and Eisermann's paper \cite{MichaelEisermann:2012be}. Nevertheless, we were still obliged to devise some proofs on our own, as discussed later.

This paper continues as follows: we start with a motivating example (\S\ref{sec:cauchy_motivating_ex}) to explain the difficulty of formal evaluation of the winding number in Isabelle/HOL. We then present an intuitive description of the link between the winding number and the Cauchy indices (\S\ref{sec:cindex_intuition}), which is then developed formally (\S\ref{sec:evaluate_winding}). Next, we present verified procedures that count the number of complex roots in a domain (\S\ref{sec:root_counting}), along with some limitations (\S\ref{sec:cindex_limitations}) and make some general remarks on the formalisation (\S\ref{sec:cindex_remarks}). Finally, we discuss related work (\S\ref{sec:related_work}) and present conclusions (\S\ref{sec:conclusion}).

\section{A Motivating Example} \label{sec:cauchy_motivating_ex}

\begin{figure*}[t!]
	\centering
	\begin{subfigure}[t]{0.5\textwidth}
		\centering
		\begin{tikzpicture}[scale=1.2]

		\draw [blue]  (2,0) arc (0:180:2) [arrow inside={end=stealth,opt={,scale=2}}{0.3,0.75}];
		\draw [dashed]  (-2,0) arc (180:360:2) [arrow inside={end=stealth,opt={,scale=2}}{0.3,0.75}];
		\draw [blue] (-2,0) -- (2,0) [arrow inside={end=stealth,opt={,scale=2}}{0.3,0.8}];

		\draw [fill] (0,1) circle [radius=.3mm] ;
		\node at (0,1) [right] {$i$};

		\node at (0,0) [below] {$L_r$};

		\draw [fill] (-2,0) circle [radius=.3mm] node [left] {$-r$} ;
		\draw [fill] (2,0) circle [radius=.3mm] node [right] {$r$} ;

		\node at (-1,-2.2) {$C'_r$};
		\node at (1,2) {$C_r$};
		\end{tikzpicture}
		\caption{}
	\end{subfigure}%
	~
	\begin{subfigure}[t]{0.5\textwidth}
		\centering
		\begin{tikzpicture}[scale=1.2]

		\draw [fill] (-2,0) circle [radius=.3mm] node [left] {$-r$} ;
		\draw [fill] (2,0) circle [radius=.3mm] node [right] {$r$} ;

		\draw [blue]  (2,0) arc (0:180:2) [arrow inside={end=stealth,opt={,scale=2}}{0.3,0.75}];
		\draw [blue]  (-2,0) -- (2,0) [arrow inside={end=stealth,opt={,scale=2}}{0.3,0.8}];

		\draw [dashed,->] (0,-1) -- (0,-3);

		\draw [fill] (0,-1) circle [radius=.3mm] ;
		\node at (0,-1) [right] {$-i$};

		\node at (0,0) [below] {$L_r$};

		\node at (0,-2) [left] {$L'_r$};

		\node at (1,2) {$C_r$};
		\end{tikzpicture}
		\caption{}
	\end{subfigure}
	\caption{Complex points $(0,-i)$ and $(0,i)$, and a closed path $L_r+C_r$}
	\label{fig:semicircle_old}
\end{figure*}
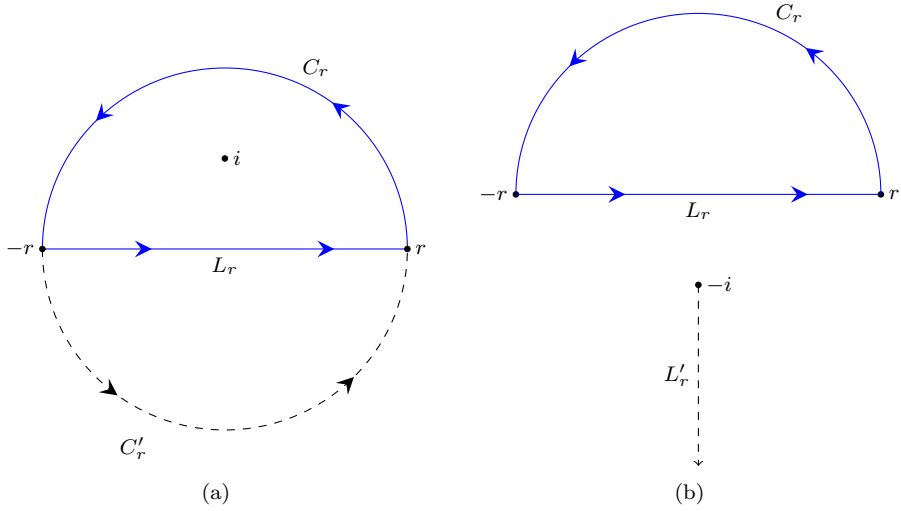

 In the formalisation of Cauchy's residue theorem \cite{Li_ITP2016}, we demonstrated an application of this theorem to formally evaluate an improper integral in Isabelle/HOL:
 \begin{equation} \label{eq:m_improper}
	\int_{-\infty}^{\infty} \frac{d x}{x^2+1} = \pi.
 \end{equation}
 The idea is to embed this integral into the complex plane, and, as illustrated in Fig.~\ref{fig:semicircle_old}, to construct a linear path $L_r$ from $-r$ to $r$ and a semi-circular path $C_r$ centred at 0 with radius $r>1$:
  \[
 C_r (t) = r e^{i \pi t} \qquad \mathrm{for}\quad t \in [0,1],
 \]
 \[
 L_r (t) = (1-t) (- r) + t r \qquad \mathrm{for}\quad t \in [0,1].
 \]
 Next, by letting
 \[
	 f(w) = \frac{1}{w^2 + 1},
 \]
 and $r \rightarrow \infty$, we can derive (\ref{eq:m_improper}) through the following steps:
 \begin{align}
	 \int_{-\infty}^{\infty} \frac{d x}{x^2+1} &= \oint_{L_r} f \\
 	& = \oint_{L_r + C_r} f  \label{eq:motivation_1} \\
 	&= n(L_r + C_r,i)\Res(f,i) + n(L_r + C_r,-i)\Res(f,-i) \label{eq:motivation_2} \\
 	& = \pi. \label{eq:motivation_3}
 \end{align}
 Here $L_r+C_r$ is formed by appending $C_r$ to the end of $L_r$, and $\Res(f,i)$ is the residue of $f$ at $i$. Equation (\ref{eq:motivation_1}) is because $\oint_{C_r} f = 0$ as $r \rightarrow \infty$. The application of the residue theorem is within (\ref{eq:motivation_2}); we exploit the fact that $i$ and $-i$ are the only two singularities of $f$ over the complex plane, since
 \[
	 \frac{1}{w^2 + 1} = \frac{1}{(w-i)(w+i)}.
 \]

While carrying out the formal proofs of (\ref{eq:motivation_3}), surprisingly, the most troublesome part of the proof is to evaluate the winding numbers:
\begin{equation} \label{eq:n_eq_1}
n(L_r+C_r,i) = 1
\end{equation}
\begin{equation} \label{eq:n_eq_0}
n(L_r+C_r,-i) = 0.
\end{equation}

Equations (\ref{eq:n_eq_1}) and (\ref{eq:n_eq_0}) are straightforward to humans, as it can be seen from Fig.~\ref{fig:semicircle_old} that $L_r+C_r$ passes counterclockwise around the point $i$  exactly one time, and around $-i$ zero times. However, formally deriving these facts was non-trivial.

\begin{example}[Proof of $n(L_r+C_r,i) = 1$] \label{ex:n_1}
	We defined an auxiliary semi-circular path $C'_r$ where
	\[
	C'_r (t) = r e^{i \pi (t+1)} \qquad \mathrm{for}\quad t \in [0, 1]
	\]
	as can be seen in Fig.~\ref{fig:semicircle_old}a. As $C_r+C'_r$ forms a (full) circular path with $i$ lying inside the circle, we had 
	\begin{equation} \label{eq:motivation_4}
		n(C_r+C'_r,i)  =1.
	\end{equation}
	In addition, we further proved that $C_r+C'_r$ and $L_r+C_r$ are homotopic on the space of the complex plane except for the point $i$ (i.e., on $\mathbb{C} - \{i\}$), and hence 
	\begin{equation} \label{eq:motivation_5}
		n(L_r+C_r,i) = n(C_r+C'_r,i)
	\end{equation}
	by using the following Isabelle lemma:
	\begin{lemma}[\isa{winding\_number\_homotopic\_paths}]
		\vspace{-7pt}
		\begin{isabellebody}
			\isanewline
			\ \ \isakeyword{fixes}\ z::complex\ \isakeyword{and}\ \isasymgamma\isactrlsub 1\ \isasymgamma\isactrlsub 2::"real\ \isasymRightarrow \ complex"\isanewline
			\ \ \isakeyword{assumes}\ "homotopic\_paths\ (-\isacharbraceleft z\isacharbraceright )\ \isasymgamma\isactrlsub 1\ \isasymgamma\isactrlsub 2"\isanewline
			\ \ \isakeyword{shows}\ "winding\_number\ \isasymgamma\isactrlsub 1\ z\ =\ winding\_number\ \isasymgamma\isactrlsub {2}\ z"\
		\end{isabellebody}
	\end{lemma}
	\noindent where \isa{winding\_number\ \isasymgamma\isactrlsub 1\ z} encodes the winding number of $\gamma_1$ around $z$: $n(\gamma_1,z)$, and \isa{homotopic\_paths} encodes the homotopic proposition between two paths.
	Putting (\ref{eq:motivation_4}) and (\ref{eq:motivation_5}) together yields $n(L_r+C_r,i) = 1$, which concludes the whole proof.
\end{example}

\begin{example}[Proof of $n(L_r+C_r,-i) = 0$] \label{ex:n_0}
	We started by defining a ray $L'_r$ starting from $-i$ and pointing towards the negative infinity of the imaginary axis:
	\[
	L'_r (t) = (- i) - t i \qquad \mathrm{for}\quad t \in [0, \infty)
	\]
	as illustrated in Fig.~\ref{fig:semicircle_old}b. Subsequently, we showed that 
	\begin{equation} \label{eq:motivation_6}
		L'_R \mbox{ does not intersect with } L_r+C_r,
	\end{equation}
	and then applied the following lemma in Isabelle
	\begin{lemma}[\isa{winding\_number\_less\_1}] \label{thm:winding_number_less_1}
		\vspace{-7pt}
		\begin{isabellebody}
			\isanewline
			\ \ \isakeyword{fixes}\ z\ w::complex\ \isakeyword{and}\ \isasymgamma ::"real\ \isasymRightarrow \ complex"\isanewline
			\ \ \isakeyword{assumes}\ "valid\_path\ \isasymgamma "\ \isakeyword{and}\ "z\ \isasymnotin \ path\_image\ \isasymgamma "\ \isakeyword{and}\ "w\ \isasymnoteq \ z"\ \isanewline
			\ \ \ \ \isakeyword{and}\ not\_intersection:"\isasymAnd a::real.\ 0\ <\ a\ \isasymLongrightarrow \ z\ +\ a*(w\ -\ z)\ \isasymnotin \ path\_image\ \isasymgamma "\isanewline
			\ \ \isakeyword{shows}\ "\isasymbar Re(winding\_number\ \isasymgamma \ z)\isasymbar \ <\ 1"
		\end{isabellebody}
	\end{lemma}
	\noindent where 
	\begin{itemize}
		\item \isa{valid\_path \isasymgamma} assumes that $\gamma$ is piecewise continuously differentiable on $[0,1]$,
		\item \isa{z\ \isasymnotin \ path\_image\ \isasymgamma} asserts that $z$ is not on the path $\gamma$,
		\item the assumption \isa{not\_intersection} asserts that the ray starting at $z \in \mathbb{C}$ and through $w \in \mathbb{C}$ ($\{z + a (w-z) \mid  a>0 \}$) does not intersect with $\gamma$---for all $a>0$, $z + a (w-z)$ does not lie on $\gamma$.
	\end{itemize}
	Note that the real part of a winding number $\mathrm{Re}(n(\gamma,z))$ measures the degree of the winding: in case of $\gamma$ winding around $z$ counterclockwise for exactly one turn, we have $n(\gamma,z) = \mathrm{Re}(n(\gamma,z)) = 1$. Essentially, Lemma \ref{thm:winding_number_less_1} claims that a path $\gamma$ can only wind around $z$ for less than one turn, $|\mathrm{Re}(n(\gamma,z))| < 1$, if there is a ray starting at $z$ and not intersecting with $\gamma$. Joining Lemma~\ref{thm:winding_number_less_1} with (\ref{eq:motivation_6}) leads to
	\begin{equation} \label{eq:motivation_7}
		|\mathrm{Re}(n(L_r+C_r,-i))|  < 1.
	\end{equation}
	Moreover, as $L_r+C_r$ is a closed path, 
	\begin{equation} \label{eq:motivation_8}
		n(L_r+C_r,-i) \in \mathbb{Z}
	\end{equation}
	By combining (\ref{eq:motivation_7}) and (\ref{eq:motivation_8}) , we managed to derive $n(L_r+C_r,-i)=0$.
\end{example}

As can be observed in Examples \ref{ex:n_1} and \ref{ex:n_0}, our proofs of  $n(L_r+C_r,i) = 1$ and $n(L_r+C_r,-i) = 0$ were ad hoc, and involved the manual construction of auxiliary paths or rays (e.g. $C'_R$ and $L'_R$). Similar difficulties have also been mentioned by John Harrison when formalising the prime number theorem \cite{harrison-pnt}. In the next section, we will introduce an idea to systematically evaluate winding numbers.

\section{The Intuition} \label{sec:cindex_intuition}

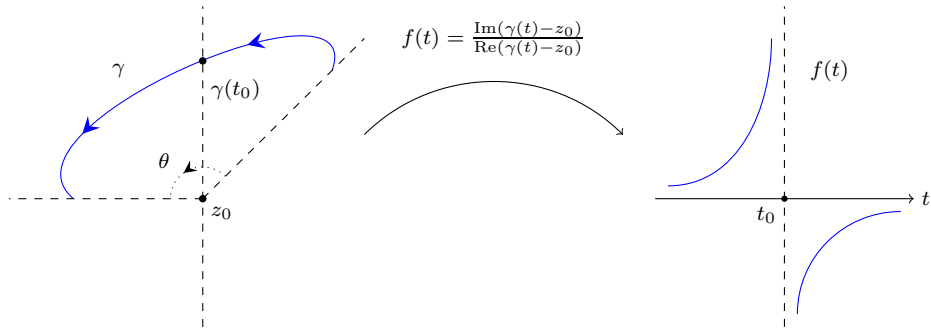
\begin{figure}[ht]
	\centering
	\begin{tikzpicture}[scale=0.85]

	\draw [blue] (2,2) to[out=70,in=140,looseness=1] (-2,0)  [arrow inside={end=stealth,opt={,scale=2}}{0.3,0.8}];

	\draw [dashed] (0,0) -- (2.5,2.5);
	\draw [dashed] (0,0) -- (-3,0);
	\draw [dotted] (.35,.35) arc (45:180:.5)  [arrow inside={end=stealth,opt={,scale=1.5}}{0.6}];


	\draw[dashed] (0,-2) -- (0,3);


	\draw [fill] (0,2.15) circle [radius=.5mm];

	\draw [fill] (0,0) circle [radius=.5mm] ;
	\node at (0,0) [below right] {$z_0$};

	\node at (-.6,.6) {$\theta$};
	\node at(-1.3,2) {$\gamma$};
	\node at (0,1.7) [right] {$\gamma(t_0)$};

	\draw[dashed] (9,-2)--(9,3) node[right]{};
	\draw[->] (7,0)--(11,0) node[right]{$t$};

	\draw [blue] (7.2,0.2) to[out=0,in=270,looseness=1] (8.8,2.5);
	\draw [blue] (9.2,-1.8) to[out=90,in=180,looseness=1] (10.8,-0.2);

	\draw [->] (2.5,1) to[out=45,in=135,looseness=1] (6.5,1);

	\node at (9.7,2) {$f(t)$};
	\node at (4.5,2.5) {$f(t) = \frac{\Im(\gamma(t) - z_0)}{\Re(\gamma(t) - z_0)}$};

	\draw [fill] (9,0) circle [radius=.4mm] ;
	\node at(9,0) [below left] {$t_0$};
	\end{tikzpicture}
	\caption{Left: a path $\gamma$ crosses the line $\{z \mid \Re(z)=\Re(z_0)\}$ at $\gamma(t_0)$ such that $\Re(\gamma(t_0)) > \Re(z_0)$. Right: the image of $f$ as a point travels through $\gamma$}
	\label{fig:simp_crossing}
\end{figure}

The fundamental idea of evaluating a winding number $n(\gamma,z_0)$ in this paper is to reduce the evaluation to \emph{classifications} of \textit{how} paths cross the line $\{z \mid \Re(z)=\Re(z_0)\}$: continuously or not and in which direction.

In a simple case, suppose a path $\gamma$ crosses the line $\{z \mid \Re(z)=\Re(z_0)\}$ exactly once at the point $\gamma(t_0)$ such that $\Im(\gamma(t_0)) > \Im(z_0)$ (see Fig. \ref{fig:simp_crossing} (left)), and let $\theta$ be the change in the argument of a complex point travelling through $\gamma$. It should not be hard to observe that
\[
0<\theta < 2 \pi,
\]
and by considering $\Re (n(\gamma,z_0)) = \theta  / (2 \pi)$ we can have
\[
0 < \Re (n(\gamma,z_0)) < 1,
\]
which is an approximation of $ \Re (n(\gamma,z_0))$. That is, we have approximated $\Re(n(\gamma,z_0))$ by the way that $\gamma$ crosses the line $\{z \mid \Re(z)=\Re(z_0)\}$.

To make this idea more precise, let
\[
f(t) = \frac{\Im(\gamma(t) - z_0)}{\Re(\gamma(t) - z_0)}.
\]
The image of $f$ as a point travels through $\gamma$ is as illustrated in Fig.~\ref{fig:simp_crossing} (right), where $f$ jumps from $+\infty$ to $-\infty$ across $t_0$.  We can then formally characterise those jumps.
\begin{definition} [Jump] \label{def:jumpF}
	For $f : \mathbb{R} \rightarrow \mathbb{R}$ and $x \in \mathbb{R}$, we define
	\[
	\jump_+(f,x) =
	\begin{dcases}
	\frac{1}{2} & \mbox{if } \lim_{u \rightarrow x^+} f(u)=+\infty,\\
	-\frac{1}{2} & \mbox{if } \lim_{u \rightarrow x^+} f(u)=-\infty,\\
	0 &  \mbox{ otherwise,}\\
	\end{dcases}
	\]
	\[
	\jump_-(f,x) =
	\begin{dcases}
	\frac{1}{2} & \mbox{if } \lim_{u \rightarrow x^-} f(u)=+\infty,\\
	-\frac{1}{2} & \mbox{if } \lim_{u \rightarrow x^-} f(u)=-\infty,\\
	0 &  \mbox{ otherwise.}\\
	\end{dcases}
	\]
\end{definition}
Specifically, we can conjecture that $\jump_+(f,t_0) - \jump_-(f,t_0)$ captures the way that $\gamma$ crosses the line $\{z \mid \Re(z)=\Re(z_0)\}$ in Fig.~\ref{fig:simp_crossing}, hence $\Re(n(\gamma,z_0))$ can be approximated using $\jump_+$ and $\jump_-$:
\[
\left | \Re(n(\gamma,z_0)) + \frac{\jump_+(f,t_0) - \jump_-(f,t_0)}{2} \right | < \frac{1}{2}.
\]

In more general cases, we can define Cauchy indices by summing up these jumps over an interval and along a path.
\begin{definition}[Cauchy index] \label{def:ind}
	For $f : \mathbb{R} \rightarrow \mathbb{R}$ and $a, b \in \mathbb{R}$, the Cauchy index of $f$ over a closed interval $[a,b]$ is defined as
	\[
	\Ind_a^b(f) = \sum_{x \in [a,b)} \jump_+(f,x) - \sum_{x \in (a,b]} \jump_-(f,x).
	\]
\end{definition}
\begin{definition}[Cauchy index along a path] \label{def:indp}
	Given a path $\gamma : [0,1] \rightarrow \mathbb{C}$ and a point $z_0 \in \mathbb{C}$, the Cauchy index along $\gamma$ about $z_0$ is defined as
	\[
	\Indp(\gamma,z_0) = \Ind_0^1(f)
	\]
	where
	\[
	f(t) = \frac{\Im(\gamma(t) - z_0)}{\Re(\gamma(t) - z_0)}.
	\]
\end{definition}
In particular, it can be checked that the Cauchy index $\Indp(\gamma,z_0)$ captures the way that $\gamma$ crosses the line $\{z \mid \Re(z)=\Re(z_0)\}$, hence leads to an approximation of $\Re(n(\gamma,z_0))$:
\[
\left| \Re(n(\gamma,z_0)) + \frac{ \Indp(\gamma,z_0) } {2} \right| < \frac{1}{2}.
\]
More interestingly, by further knowing that $\gamma$ is a loop we can derive $\Re(n(\gamma,z_0)) = n(\gamma,z_0) \in \mathbb{Z}$ and $\Indp(\gamma,z_0) / 2 \in \mathbb{Z}$, following which we come to the core proposition of this paper:
\begin{proposition} \label{prop:winding_eq}
	Given a valid path $\gamma : [0,1] \rightarrow \mathbb{C}$ and a point $z_0 \in \mathbb{C}$, such that $\gamma$ is a loop and $z_0$ is not on the image of $\gamma$, we have
	\[
	n(\gamma,z_0) = - \frac{\Indp(\gamma,z_0)}{2}.
	\]
\end{proposition}
\noindent That is, under some assumptions, we can evaluate a winding number through Cauchy indices!

\begin{figure*}[ht]
	\centering
	\begin{subfigure}[t]{0.5\textwidth}
		\centering
		\begin{tikzpicture}[scale=1.2]

		\draw[dashed] (0,-2) -- (0,3);

		\draw [blue]  (2,0) arc (0:180:2) [arrow inside={end=stealth,opt={,scale=2}}{0.3,0.75}];
		\draw [blue]  (-2,0) -- (2,0) [arrow inside={end=stealth,opt={,scale=2}}{0.3,0.8}];

		\draw [fill] (0,1) circle [radius=.4mm] ;
		\node at (0,1) [right] {$i$};

		\node at (.5,0) [below] {$L_r$};

		\draw [fill] (-2,0) circle [radius=.3mm] node [left] {$-r$} ;
		\draw [fill] (2,0) circle [radius=.3mm] node [right] {$r$} ;





		\node at (1,2) {$C_r$};
		\end{tikzpicture}
		\caption{}
	\end{subfigure}%
	~
	\begin{subfigure}[t]{0.5\textwidth}
		\centering
		\begin{tikzpicture}[scale=1.2]

		\draw[dashed] (0,-2) -- (0,3);

		\draw [blue]  (2,0) arc (0:180:2) [arrow inside={end=stealth,opt={,scale=2}}{0.3,0.75}];
		\draw [blue]  (-2,0) -- (2,0) [arrow inside={end=stealth,opt={,scale=2}}{0.3,0.8}];


		\draw [fill] (0,-1) circle [radius=.4mm] ;
		\node at (0,-1) [right] {$-i$};

		\draw [fill] (-2,0) circle [radius=.3mm] node [left] {$-r$} ;
		\draw [fill] (2,0) circle [radius=.3mm] node [right] {$r$} ;


		\node at (.5,0) [below] {$L_r$};


		\node at (1,2) {$C_r$};
		\end{tikzpicture}
		\caption{}
	\end{subfigure}
	\caption{Evaluating $n(L_r+C_r,i)$ and $n(L_r+C_r,-i)$ through the way that the path $L_r+C_r$ crosses the imaginary axis}
	\label{fig:semicircle_new}
\end{figure*}
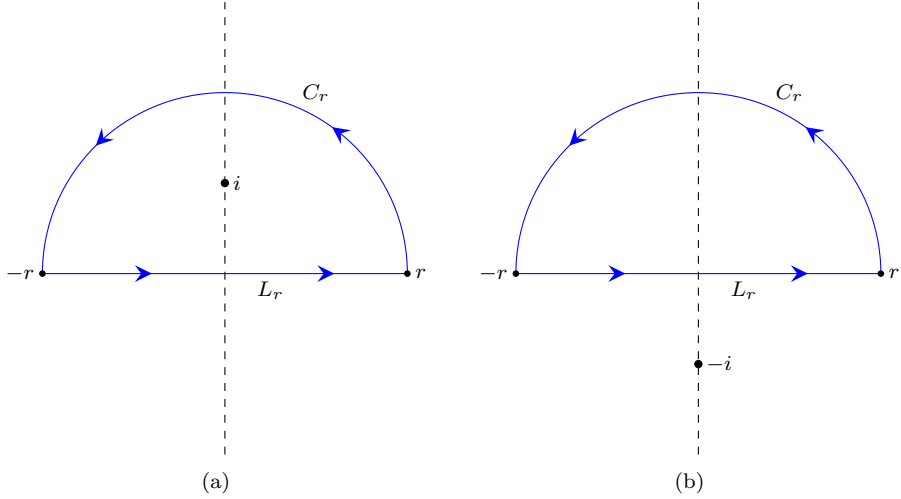

A formal proof of Proposition~\ref{prop:winding_eq} will be introduced in \S\ref{sec:formal_proof_prop}. Here, given the statement of the proposition, we can have alternative proofs for $n(L_r+C_r,i) = 1$ and $n(L_r+C_r,-i) = 0$.

\begin{example}[Alternative proof of $n(L_r+C_r,i) = 1$] \label{ex:n_1_alt}
	As $L_r+C_r$ is a loop, applying Proposition \ref{prop:winding_eq} yields
	\[
	n(L_r+C_r,i) = - \frac{\Indp(L_r+C_r,i)}{2} = -\frac{1}{2} (\Indp(L_r,i) + \Indp(C_r,i)),
	\]
	which reduces $n(L_r+C_r,i)$ to the evaluations of $\Indp(L_r,i)$ and $\Indp(C_r,i)$. In this case, by definition we can easily decide $\Indp(L_r,i) = -1$ and $\Indp(C_r,i) = -1$ as illustrated in Fig.~\ref{fig:semicircle_new}a. Hence, we have
	\[
	n(L_r+C_r,i) =  -\frac{1}{2} ((-1) + (-1)) = 1
	\]
	and conclude the proof.
\end{example}
\begin{example}[Alternative proof of $n(L_r+C_r,-i) = 0$] \label{ex:n_0_alt}
	As shown in Fig.~\ref{fig:semicircle_new}b, we can similarly have
	\[
	\begin{split}
	n(L_R+C_R,-i) &= - \frac{\Indp(L_r+C_r,-i)}{2} \\
	&= -\frac{1}{2} (\Indp(L_r,-i) + \Indp(C_r,-i))\\
	&=  -\frac{1}{2} (1 + (-1)) = 0
	\end{split}
	\]
	by which the proof is completed.
\end{example}

Compared to the previous proofs presented in Examples \ref{ex:n_1} and \ref{ex:n_0}, the alternative proofs in Examples \ref{ex:n_1_alt} and \ref{ex:n_0_alt} are systematic and less demanding to devise once we have a formalisation of Proposition  \ref{prop:winding_eq}, which is what we will introduce in the next section.

\section{Evaluating Winding Numbers} \label{sec:evaluate_winding}
The previous section presented an informal intuition to systematically evaluate winding numbers; in this section, we will report the formal development of this intuition. We will first present a mechanised proof of Proposition \ref{prop:winding_eq} (\S\ref{sec:formal_proof_prop}), which includes mechanised definitions of jumps and Cauchy indices (i.e., Definition \ref{def:jumpF}, \ref{def:ind} and \ref{def:indp}) and several related properties of these objects. After that, we build a tactic in Isabelle/HOL that is used to mechanise proofs presented in Example \ref{ex:n_1_alt} and \ref{ex:n_0_alt} (\S\ref{sec:tactic_for_evaluation}). Finally, we discuss some subtleties we encountered during the formalisation (\S\ref{sec:subtleties}).

\subsection{A Formal Proof of Proposition \ref{prop:winding_eq}} \label{sec:formal_proof_prop}
For $\jump_-$ and $\jump_+$ (see Definition \ref{def:jumpF}), we have used the filter mechanism \cite{hoelzl2013typeclasses} to define a function \isa{jumpF}:
\begin{isabelle}
	\isacommand{definition}	\ jumpF::"(real\ \isasymRightarrow \ real)\ \isasymRightarrow \ real\ filter\ \isasymRightarrow \ real"\ \isakeyword{where}\ \isanewline
	\ \ "jumpF\ f\ F\ \isasymequiv \ (if\ (LIM\ x\ F.\ f\ x\ :>\ at\_top)\ then\ 1/2\ else\ \isanewline
	\ \ \ \ \ \ \ \ \ \ \ \ \ \ \ \ \ \ if\ (LIM\ x\ F.\ f\ x\ :>\ at\_bot)\ then\ -1/2\ else\ 0)"
\end{isabelle}
and encoded $\jump_-(f,x)$ and $\jump_+(f,x)$ as 
\begin{quote}
	\isa{jumpF f (at\_left x)} and  \isa{jumpF f (at\_right x)},
\end{quote}
 respectively. Here, \isa{at\_left x}, \isa{at\_right x}, \isa{at\_top}, and \isa{at\_bot} are all filters, where a filter is a predicate on predicates that satisfies certain properties. Filters are extensively used in the analysis library of Isabelle to encode varieties of logical quantification: for example,  \isa{at\_left x} encodes the statement ``for a variable that is sufficiently close to $x$ from the left", and \isa{at\_top} represents ``for a sufficiently large variable". Furthermore, \isa{LIM\ x\ (at\_left\ x).\ f\ x\ :>\ at\_top} encoded the proposition
\begin{equation} \label{eq:formal_proof_prop_1}
		\lim_{u \rightarrow x^-} f(u) = +\infty,
\end{equation}
and this encoding can be justified by the following equality in Isabelle:
\begin{isabelle}
	(LIM\ x\ (at\_left\ x).\ f\ x\ :>\ at\_top)\ =\ (\isasymforall z.\ \isasymexists b<x.\ \isasymforall y>b.\ y\ <\ x\ \isasymlongrightarrow \ z\ \isasymle \ f\ y)
\end{isabelle}
where \isa{\isasymforall z.\ \isasymexists b<x.\ \isasymforall y>b.\ y\ <\ x\ \isasymlongrightarrow \ z\ \isasymle \ f\ y} matches the usual definition of (\ref{eq:formal_proof_prop_1}) in textbooks.

We can then encode $\Ind_a^b(f)$ and $\Indp(\gamma,z_0)$ (see Definitions \ref{def:ind} and \ref{def:indp}) as \isa{cindexE} and \isa{cindex\_pathE} respectively:
\begin{isabelle}
	\isacommand{definition}	\ cindexE::"real\ \isasymRightarrow \ real\ \isasymRightarrow \ (real\ \isasymRightarrow \ real)\ \isasymRightarrow \ real"\ \isakeyword{where}\isanewline
	\ \ "cindexE\ a\ b\ f\ =\ \isanewline
	\ \ \ \ \ \ (\isasymSum x\isasymin \isacharbraceleft x.\ jumpF\ f\ (at\_right\ x)\ \isasymnoteq\ 0\ \isasymand \ a\ \isasymle\ x\ \isasymand \ x\ <\ b\isacharbraceright .\ jumpF\ f\ (at\_right\ x))\isanewline
	\ \ \ \ \ \ -\ (\isasymSum x\isasymin \isacharbraceleft x.\ jumpF\ f\ (at\_left\ x)\ \isasymnoteq\ 0\ \isasymand \ a\ <\ x\ \isasymand \ x\ \isasymle\ b\isacharbraceright .\ jumpF\ f\ (at\_left\ x))"
\end{isabelle}
\begin{isabelle}
	\isacommand{definition}	\ cindex\_pathE::"(real\ \isasymRightarrow \ complex)\ \isasymRightarrow \ complex\ \isasymRightarrow \ real"\ \isakeyword{where}\isanewline
	\ \ "cindex\_pathE\ \isasymgamma \ z\isactrlsub 0\ =\ cindexE\ 0\ 1\ (\isasymlambda t.\ Im\ (\isasymgamma \ t\ -\ z\isactrlsub 0)\ /\ Re\ (\isasymgamma \ t\ -\ z\isactrlsub 0))"
\end{isabelle}
Note, in the definition of $\Ind_a^b(f)$ we have a term
\[
\sum_{x \in [a,b)} \jump_+(f,x)
\]
which actually hides an assumption: that only a finite number of points within the interval $[a,b)$ contribute to the sum. This assumption is made explicit when \isa{cindexE} is defined by summing jumps over the following set:
\begin{isabelle}
	\ \ \ \ \ \ \ \ \ \ \ \ \ \ \ \isacharbraceleft x.\ jumpF\ f\ (at\_right\ x)\ \isasymnoteq \ 0\ \isasymand \ a\ \isasymle \ x\ \isasymand \ x\ <\ b\isacharbraceright .
\end{isabelle}
If the set above is infinite (i.e., the sum $\sum_{x \in [a,b)} \jump_+(f,x)$ is not mathematically well-defined) we have
\begin{isabelle}
	(\isasymSum x\isasymin \isacharbraceleft x.\ jumpF\ f\ (at\_right\ x)\ \isasymnoteq\ 0\ \isasymand \ a\ \isasymle\ x\ \isasymand \ x\ <\ b\isacharbraceright .\ jumpF\ f\ (at\_right\ x))\ =\ 0.
\end{isabelle}
In other words, Isabelle/HOL deems the sum over an infinite set to denote zero. 

Due to the issue of well-defined sums, many of our lemmas related to \isa{cindexE} should assume a finite number of jumps:
\begin{isabelle}
	\isacommand{definition}	\ finite\_jumpFs::"(real\ \isasymRightarrow \ real)\ \isasymRightarrow \ real\ \isasymRightarrow \ real\ \isasymRightarrow \ bool"\ \isakeyword{where}\isanewline
	\ \ "finite\_jumpFs\ f\ a\ b\ =\ finite\ \isacharbraceleft x.\ (jumpF\ f\ (at\_left\ x)\ \isasymnoteq\ 0\ \isanewline
	\ \ \ \ \ \ \ \ \ \ \ \ \ \ \ \ \ \ \ \ \ \ \ \ \ \ \ \ \isasymor \ jumpF\ f\ (at\_right\ x)\ \isasymnoteq\ 0)\ \isasymand \ a\ \isasymle\ x\ \isasymand \ x\ \isasymle\ b\isacharbraceright "
\end{isabelle}
which guarantees the well-definedness of \isa{cindexE}.

Now, suppose that we know that $\Indp$ is well-defined: there are only a finite number of jumps over the path. What strategy can we employ to formally prove Proposition \ref{prop:winding_eq}? Naturally,
we may want to divide the path into a finite number of segments (subpaths) separated by those jumps, and then perform inductions on these segments. To formalise the finiteness of such segments, we defined an inductive predicate:
\begin{isabelle}
	\isacommand{inductive}	\ finite\_Psegments::"(real\ \isasymRightarrow \ bool)\ \isasymRightarrow \ real\ \isasymRightarrow \ real\ \isasymRightarrow \ bool"\ \isanewline
	\ \ \ \ \isakeyword{for}\ P\ \isakeyword{where}\isanewline
	\ \ emptyI:\ "a\isasymge b\ \isasymLongrightarrow \ finite\_Psegments\ P\ a\ b"|\isanewline
	\ \ insertI\_1:\ "\isasymlbrakk s\isasymin \isacharbraceleft a..<b\isacharbraceright ;\ s=a\ \isasymor \ P\ s;\ \isasymforall t\isasymin \isacharbraceleft s<..<b\isacharbraceright .\ P\ t;\ \isanewline
	\ \ \ \ \ \ finite\_Psegments\ P\ a\ s\isasymrbrakk \ \isasymLongrightarrow \ finite\_Psegments\ P\ a\ b"|\isanewline
	\ \ insertI\_2:\ "\isasymlbrakk s\isasymin \isacharbraceleft a..<b\isacharbraceright ;\ s=a\ \isasymor \ P\ s;\ \isasymforall t\isasymin \isacharbraceleft s<..<b\isacharbraceright .\ \isasymnot P\ t;\isanewline
	\ \ \ \ \ \ finite\_Psegments\ P\ a\ s\isasymrbrakk \ \isasymLongrightarrow \ finite\_Psegments\ P\ a\ b"
\end{isabelle}

\begin{isabelle}
	\isacommand{definition}	\ finite\_ReZ\_segments::"(real\ \isasymRightarrow \ complex)\ \isasymRightarrow \ complex\ \isasymRightarrow \ bool"\ \isakeyword{where}\isanewline
	\ \ "finite\_ReZ\_segments\ \isasymgamma \ z\isactrlsub 0\ =\ finite\_Psegments\ (\isasymlambda t.\ Re\ (\isasymgamma \ t\ -\ z\isactrlsub 0)\ =\ 0)\ 0\ 1"
\end{isabelle}
The idea behind \isa{finite\_ReZ\_segments}	is that a jump of
\[
f(t) = \frac{\Im(\gamma(t) - z_0)}{\Re(\gamma(t) - z_0)}
\]
takes place only if $\lambda t.\, \Re(\gamma(t) - z_0)$ changes from 0 to $\neq 0$ (or vice versa). Hence, each of the segments of the path $\gamma$ separated by those jumps has either $\lambda t.\, \Re(\gamma(t) - z_0) = 0$ or $\lambda t.\, \Re(\gamma(t) - z_0) \neq 0$.

As can be expected, the finiteness of jumps over a path can be derived by the finiteness of segments:
\begin{lemma} [\isa{finite\_ReZ\_segments\_imp\_jumpFs}] \label{thm:finite_ReZ_segments_imp_jumpFs}
	\vspace{-7pt}
	\begin{isabellebody}
		\isanewline
		\ \ \isakeyword{fixes}\ \isasymgamma ::"real\ \isasymRightarrow \ complex"\ \isakeyword{and}\ z\isactrlsub 0::complex\isanewline
		\ \ \isakeyword{assumes}\ "finite\_ReZ\_segments\ \isasymgamma \ z\isactrlsub 0"\ \isakeyword{and}\ "path\ \isasymgamma "\ \isanewline
		\ \ \isakeyword{shows}\ "finite\_jumpFs\ (\isasymlambda t.\ Im\ (\isasymgamma \ t\ -\ z\isactrlsub 0)/Re\ (\isasymgamma \ t\ -\ z\isactrlsub 0))\ 0\ 1"
	\end{isabellebody}
\end{lemma}
\noindent
where \isa{path\ \isasymgamma} asserts that $\gamma$ is a continuous function on $[0..1]$ (so that it is a path). Roughly speaking, Lemma \ref{thm:finite_ReZ_segments_imp_jumpFs} claims that a path will have a finite number of jumps if it can be divided into a finite number of segments.

By assuming such a finite number of segments we have well-defined \isa{cindex\_pathE}, and can then derive some useful related properties:
\begin{lemma} [\isa{cindex\_pathE\_subpath\_combine}] \label{thm:cindex_pathE_subpath_combine}
		\vspace{-7pt}
	\begin{isabellebody}
		\isanewline
		\ \ \isakeyword{fixes}\ \isasymgamma ::"real\ \isasymRightarrow \ complex"\ \isakeyword{and}\ z\isactrlsub 0::complex\isanewline
		\ \ \isakeyword{assumes}\ "finite\_ReZ\_segments\ \isasymgamma \ z\isactrlsub 0"\isakeyword{and}\ "path\ \isasymgamma "\ \isanewline
		\ \ \ \ \isakeyword{and}\ "0\isasymle a"\ \isakeyword{and}\ "a\isasymle b"\ \isakeyword{and}\ "b\isasymle c"\ \isakeyword{and}\ "c\isasymle 1"\isanewline
		\ \ \isakeyword{shows}\ "cindex\_pathE\ (subpath\ a\ b\ \isasymgamma )\ z\isactrlsub 0\ +\ cindex\_pathE\ (subpath\ b\ c\ \isasymgamma )\ z\isactrlsub 0\ \isanewline
		\ \ \ \ \ \ \ \ \ \ =\ cindex\_pathE\ (subpath\ a\ c\ \isasymgamma )\ z\isactrlsub 0"
	\end{isabellebody}
\end{lemma}
\noindent
where \isa{subpath\ a\ b\ \isasymgamma} gives a sub-path of $\gamma$ based on parameters $a$ and $b$:
\begin{isabelle}
	\isacommand{definition}	\ subpath\ ::\ "real\ \isasymRightarrow \ real\ \isasymRightarrow \ (real\ \isasymRightarrow \ 'a)\ \isasymRightarrow \ real\ \isanewline
	\ \ \ \ \isasymRightarrow \ 'a::real\_normed\_vector"\isanewline
	\ \ \isakeyword{where}\ "subpath\ a\ b\ \isasymgamma \ \isasymequiv \ (\isasymlambda t.\ \isasymgamma ((b\ -\ a)\ *\ t\ +\ a))"
\end{isabelle}
Essentially, Lemma \ref{thm:cindex_pathE_subpath_combine} indicates that we can combine Cauchy indices along consecutive parts of a path: given a path $\gamma$ and three parameters $a,b, c$ with $0 \leq a \leq b \leq c \leq 1$, we have
\[
	\Indp(\gamma_1,z_0) + \Indp(\gamma_2,z_0) = \Indp(\gamma_3,z_0).
\]
where $\gamma_1 = \lambda t.\, \gamma((b-a) t + a)$, $\gamma_2 = \lambda t.\, \gamma((c-b) t + b)$ and $\gamma_3 = \lambda t.\, \gamma((c-a) t + a)$.

More importantly, we now have an induction rule for a path with a finite number of segments:
\begin{lemma} [\isa{finite\_ReZ\_segments\_induct}] \label{thm:finite_ReZ_segments_induct}
		\vspace{-7pt}
		\begin{isabellebody}
			\isanewline
			\ \ \isakeyword{fixes}\ \isasymgamma ::"real\ \isasymRightarrow \ complex"\ \isakeyword{and}\ z\isactrlsub 0::complex\isanewline
			\ \ \ \ \isakeyword{and}\ P::"(real\ \isasymRightarrow \ complex)\ \isasymRightarrow \ complex\ \isasymRightarrow \ bool"\isanewline
			\ \ \isakeyword{assumes}\ "finite\_ReZ\_segments\ \isasymgamma \ z\isactrlsub 0"\isanewline
			\ \ \ \ \isakeyword{and}\ sub0:"\isasymAnd g\ z.\ (P\ (subpath\ 0\ 0\ g)\ z)"\ \isanewline
			\ \ \ \ \isakeyword{and}\ subEq:"(\isasymAnd s\ g\ z.\ \isasymlbrakk s\ \isasymin \ \isacharbraceleft 0..<1\isacharbraceright ;\ s=0\ \isasymor \ Re\ (g\ s)\ =\ Re\ z;\isanewline
			\ \ \ \ \ \ \ \ \ \ \isasymforall t\ \isasymin \ \isacharbraceleft s<..<1\isacharbraceright .\ Re\ (g\ t)\ =\ Re\ z;\ \isanewline
			\ \ \ \ \ \ \ \ \ \ finite\_ReZ\_segments\ (subpath\ 0\ s\ g)\ z;\ \isanewline
			\ \ \ \ \ \ \ \ \ \ P\ (subpath\ 0\ s\ g)\ z\isasymrbrakk \ \isasymLongrightarrow \ P\ g\ z)"\ \isanewline
			\ \ \ \ \isakeyword{and}\ subNEq:"(\isasymAnd s\ g\ z.\ \isasymlbrakk s\ \isasymin \ \isacharbraceleft 0..<1\isacharbraceright ;\ s=0\ \isasymor \ Re\ (g\ s)\ =\ Re\ z;\isanewline
			\ \ \ \ \ \ \ \ \ \ \isasymforall t\ \isasymin \ \isacharbraceleft s<..<1\isacharbraceright .\ Re\ (g\ t)\ \isasymnoteq \ Re\ z;\isanewline
			\ \ \ \ \ \ \ \ \ \ finite\_ReZ\_segments\ (subpath\ 0\ s\ g)\ z;\isanewline
			\ \ \ \ \ \ \ \ \ \ P\ (subpath\ 0\ s\ g)\ z\isasymrbrakk \ \isasymLongrightarrow \ P\ g\ z)"\isanewline
			\ \ \isakeyword{shows}\ "P\ \isasymgamma \ z\isactrlsub 0"
		\end{isabellebody}
\end{lemma}
\noindent
where \isa{P} is a predicate that takes a path \isa{\isasymgamma} and a complex point \isa{z\isactrlsub {\isadigit{0}}}, and
\begin{itemize}
	\item \isa{sub0} is the base case that \isa{P} holds for a constant path;
	\item \isa{subEq} is the inductive case when the last segment is right on the line $\{x \mid \Re(x) = \Re(z) \}$: $\forall t \in (s,1).\, \Re(g(t)) = \Re(z)$;
	\item \isa{subNEq} is the inductive case when the last segment does not cross the line $\{x \mid \Re(x) = \Re(z) \}$: $\forall t \in (s,1).\, \Re(g(t)) \neq \Re(z)$.
\end{itemize}
Given a path $\gamma$ with a finite number of segments, a complex point $z_0$ and a predicate $P$ that takes a path and a complex number and returns a boolean, Lemma \ref{thm:finite_ReZ_segments_induct} provides us with an inductive rule to derive $P(\gamma,z_0)$ by recursively examining the last segment.

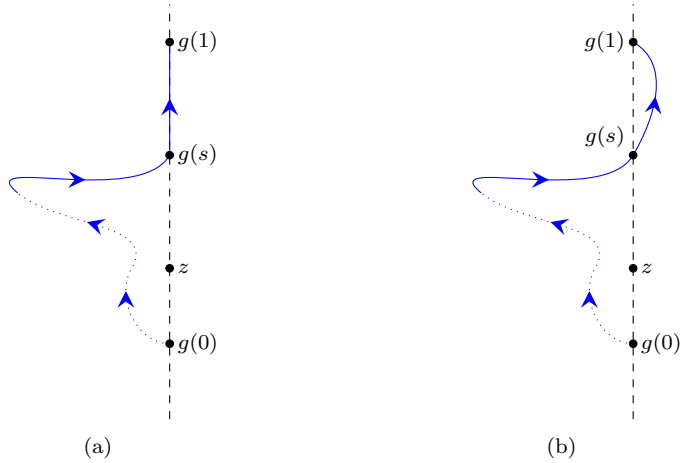
\begin{figure}[ht]
	\centering
	\begin{subfigure}[t]{0.5\textwidth}
		\centering
		\begin{tikzpicture}[scale=1]
		\draw[dashed] (0,-2)--(0,3.5) node[right]{};

		\draw [blue](0,1.5) -- (0,3)  [arrow inside={end=stealth,opt={,scale=2}}{0.5}];

		\draw [blue] (-2,1) to[out=140,in=240,looseness=1] (0,1.5) [arrow inside={end=stealth,opt={,scale=2}}{0.5}];

		[arrow inside={end=stealth,opt={,scale=2}}{.5}];

		\draw [blue,dotted] (0,-1) to[out=180,in=240,looseness=1] (-0.5,0) to[out=60,in=320,looseness=1] (-2,1) [arrow inside={end=stealth,opt={,scale=2}}{.3,.7}];

		\draw [fill] (0,3) circle [radius=.5mm] ;
		\node at (0,3) [right] {$g(1)$};

		\draw [fill] (0,-1) circle [radius=.5mm] ;
		\node at (0,-1) [right] {$g(0)$};

		\draw [fill] (0,1.5) circle [radius=.5mm] ;
		\node at (0,1.5) [right] {$g(s)$};

		\draw [fill] (0,0) circle [radius=.5mm] ;
		\node at (0,0) [right] {$z$};

		\end{tikzpicture}
		\caption{}
	\end{subfigure}%
	\hfill
	\begin{subfigure}[t]{0.5\textwidth}
		\centering
		\begin{tikzpicture}[scale=1]
		\draw[dashed] (0,-2)--(0,3.5) node[right]{};


		\draw [blue] (0,1.5) to[out=60,in=330,looseness=1] (0,3) [arrow inside={end=stealth,opt={,scale=2}}{0.5}];

		\draw [blue] (-2,1) to[out=140,in=240,looseness=1] (0,1.5) [arrow inside={end=stealth,opt={,scale=2}}{0.5}];

		[arrow inside={end=stealth,opt={,scale=2}}{.5}];

		\draw [blue,dotted] (0,-1) to[out=180,in=240,looseness=1] (-0.5,0) to[out=60,in=320,looseness=1] (-2,1) [arrow inside={end=stealth,opt={,scale=2}}{.3,.7}];

		\draw [fill] (0,3) circle [radius=.5mm] ;
		\node at (0,3) [left] {$g(1)$};

		\draw [fill] (0,-1) circle [radius=.5mm] ;
		\node at (0,-1) [right] {$g(0)$};

		\draw [fill] (0,1.5) circle [radius=.5mm] ;
		\node at (0,1.5) [above left] {$g(s)$};

		\draw [fill] (0,0) circle [radius=.5mm] ;
		\node at (0,0) [right] {$z$};

		\end{tikzpicture}
		\caption{}
	\end{subfigure}
	\caption{Inductive cases when applying Lemma \ref{thm:finite_ReZ_segments_induct}}
	\label{fig:winding_cindex_induct}
\end{figure}

Before attacking Proposition \ref{prop:winding_eq}, we can show an auxiliary lemma about $\Re(n(\gamma,z_0))$ and $\Indp(\gamma,z_0)$ when the end points of $\gamma$ are on the line $\{z \mid \Re(z) = \Re(z_0) \}$:
\begin{lemma}[\isa{winding\_number\_cindex\_pathE\_aux}]
	\label{thm:winding_number_cindex_pathE_aux}
	\vspace{-7pt}
	\begin{isabellebody}
		\isanewline
		\ \ \isakeyword{fixes}\ \isasymgamma ::"real\ \isasymRightarrow \ complex"\ \isakeyword{and}\ z\isactrlsub 0\ ::\ complex\isanewline
		\ \ \isakeyword{assumes}\ "finite\_ReZ\_segments\ \isasymgamma \ z\isactrlsub 0"\ \isakeyword{and}\ "valid\_path\ \isasymgamma "\ \isanewline
		\ \ \ \ \isakeyword{and}\ "z\isactrlsub 0\ \isasymnotin \ path\_image\ \isasymgamma "\ \isakeyword{and}\ "Re\ (\isasymgamma \ 1)\ =\ Re\ z\isactrlsub 0"\ \isanewline
		\ \ \ \ \isakeyword{and}\ "Re\ (\isasymgamma \ 0)\ =\ Re\ z\isactrlsub 0"\isanewline
		\ \ \isakeyword{shows}\ "2\ *\ Re(winding\_number\ \isasymgamma \ z\isactrlsub 0)\ =\ -\ cindex\_pathE\ \isasymgamma \ z\isactrlsub 0"
	\end{isabellebody}
\end{lemma}
\noindent Here, Lemma \ref{thm:winding_number_cindex_pathE_aux} is almost equivalent to Proposition \ref{prop:winding_eq} except for that more restrictions haven been placed on the end points of $\gamma$.

\begin{proof}[Proof of Lemma \ref{thm:winding_number_cindex_pathE_aux}]
	As there are a finite number of segments along $\gamma$ (i.e., \isa{finite{\isacharunderscore}ReZ{\isacharunderscore}segments\ {\isasymgamma}\ z\isactrlsub {\isadigit{0}}}), by inducting on these segments with Lemma \ref{thm:finite_ReZ_segments_induct} we end up with three cases. The base case is straightforward: given a constant path $g : [0,1] \rightarrow \mathbb{C}$ and a complex point $z \in \mathbb{C}$, we have $\Re(n(g,z)) = 0$ and $\Indp(g,z) = 0$,
	hence $2 \Re(n(g,z)) = - \Indp(g,z)$.

	For the inductive case when the last segment is right on the line $\{x \mid \Re(x) = \Re(z) \}$, there is $\forall t \in (s,1).\, \Re(g(t)) = \Re(z)$ as illustrated in Fig. \ref{fig:winding_cindex_induct}a. Let
	\[
	g_1(t) = g (s t)
	\]
	\[
	g_2(t) = g ((1-s) t).
	\]
	We have
	\begin{equation} \label{eq:winding_aux_1}
	n(g,z) = n(g_1,z) + n(g_2,z),
	\end{equation}
	and, by the induction hypothesis,
	\begin{equation} \label{eq:winding_aux_2}
	2 \Re(n(g_1,z)) = - \Indp(g_1,z).
	\end{equation}
	Moreover, it is possible to derive
	\begin{equation} \label{eq:winding_aux_3}
	2 \Re(n(g_2,z)) = - \Indp(g_2,z),
	\end{equation}
	since $n(g_2,z) = 0$ and $\Indp(g_2,z) = 0$.
	Furthermore, by Lemma \ref{thm:cindex_pathE_subpath_combine} we can sum up the Cauchy index along $g_1$ and $g_2$:
	\begin{equation} \label{eq:winding_aux_4}
	\Indp(g_1,z) + \Indp(g_2,z) = \Indp(g,z)
	\end{equation}
	Combining Equations (\ref{eq:winding_aux_1}), (\ref{eq:winding_aux_2}), (\ref{eq:winding_aux_3}) and (\ref{eq:winding_aux_4}) yields
	\begin{equation} \label{eq:winding_aux_5}
	\begin{split}
	2 \Re(n(g,z)) &= 2 (\Re(n(g_1,z)) + \Re(n(g_2,z))) \\
	&=  - \Indp(g_1,z)  - \Indp(g_2,z)\\
	& = - \Indp(g,z)
	\end{split}
	\end{equation}
	which concludes the case.

	For the other inductive case when the last segment does not cross the line $\{x \mid \Re(x) = \Re(z) \}$, without loss of generality, we assume
	\begin{equation} \label{eq:winding_aux_6}
	\forall t \in (s,1).\, \Re(g(t)) > \Re(z),
	\end{equation}
	and the shape of $g$ is as illustrated in Fig.~\ref{fig:winding_cindex_induct}b. Similar to the previous case, by letting $g_1(t) = g (s t)$ and $g_2(t) = g ((1-s) t)$, we have $n(g,z) = n(g_1,z) + n(g_2,z)$ and, by the induction hypothesis, $2 \Re(n(g_1,z)) = - \Indp(g_1,z)$. Moreover, by observing the shape of $g_2$ we have
	\begin{equation} \label{eq:winding_aux_7}
	2 \Re(n(g_2,z)) = \jump_-(f, 1) -\jump_+(f,0)
	\end{equation}
	\begin{equation}  \label{eq:winding_aux_8}
	\Indp(g_2,z) = \jump_+(f, 0) - \jump_-(f,1)
	\end{equation}
	where $f(t) = {\Im(g_2(t) - z)}/{\Re(g_2(t) - z)}$. Combining (\ref{eq:winding_aux_7}) with (\ref{eq:winding_aux_8}) leads to $2 \Re(n(g_2,z)) = - \Indp(g_2,z)$, following which we finish the case by deriving $2 \Re(n(g,z)) = - \Indp(g,z)$ in a way analogous to (\ref{eq:winding_aux_5}).
\end{proof}

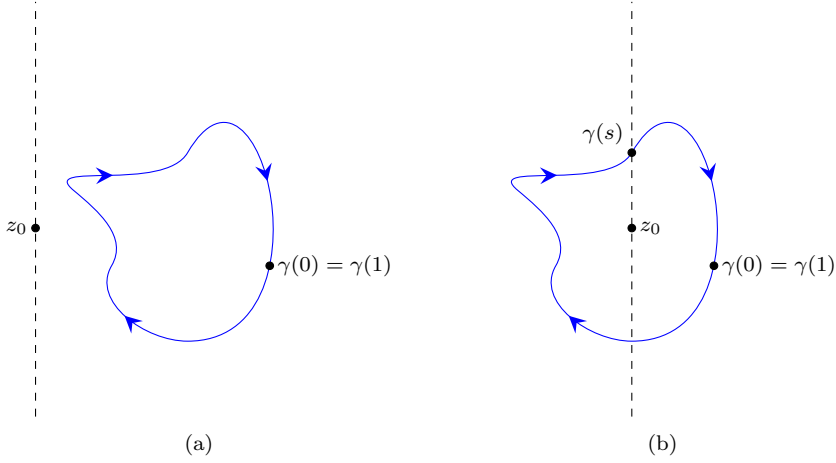
\begin{figure}[ht]
	\centering
	\begin{subfigure}[t]{0.5\textwidth}
		\centering
		\begin{tikzpicture}[scale=1]
		\draw[dashed] (-2,-2)--(-2,3.5) node[right]{};

		\draw [blue] (0,-1) to[out=180,in=240,looseness=1] (-1,0) to[out=60,in=320,looseness=1] (-1.5,1)
		to[out=140,in=240,looseness=1] (0,1.5) to[out=60,in=0,looseness=2] (0,-1) [arrow inside={end=stealth,opt={,scale=2}}{.1,.4,.7}];

		\draw [fill] (1.08,0) circle [radius=.5mm] node [right] {$\gamma(0) = \gamma(1)$} ;


		\draw [fill] (-2,.5) circle [radius=.5mm] node [left] {$z_0$};

		\end{tikzpicture}
		\caption{}
	\end{subfigure}%
	\hfill
	\begin{subfigure}[t]{0.5\textwidth}
		\centering
		\begin{tikzpicture}[scale=1]
		\draw[dashed] (0,-2)--(0,3.5) node[right]{};

		\draw [blue] (0,-1) to[out=180,in=240,looseness=1] (-1,0) to[out=60,in=320,looseness=1] (-1.5,1)
		to[out=140,in=240,looseness=1] (0,1.5) to[out=60,in=0,looseness=2] (0,-1) [arrow inside={end=stealth,opt={,scale=2}}{.1,.4,.7}];

		\draw [fill] (1.08,0) circle [radius=.5mm] node [right] {$\gamma(0) = \gamma(1)$} ;

		\draw [fill] (0,1.5) circle [radius=.5mm] node [above left] {$\gamma(s)$};

		\draw [fill] (0,.5) circle [radius=.5mm] node [right] {$z_0$};

		\end{tikzpicture}
		\caption{}
	\end{subfigure}
	\caption{To derive $n(\gamma,z_0) = - \frac{\Indp(\gamma,z_0)}{2}$ when $\gamma$ is a loop}
	\label{fig:winding_cindex_path}
\end{figure}

Finally, we are ready to formally derive Proposition \ref{prop:winding_eq} in Isabelle/HOL:
\begin{theorem}[\isa{winding\_number\_cindex\_pathE}] \label{thm:winding_number_cindex_pathE}
	\vspace{-7pt}
	\begin{isabellebody}
		\isanewline
		\ \ \isakeyword{fixes}\ \isasymgamma ::"real\ \isasymRightarrow \ complex"\ \isakeyword{and}\ z\isactrlsub 0::complex\isanewline
		\ \ \isakeyword{assumes}\ "finite\_ReZ\_segments\ \isasymgamma \ z\isactrlsub 0"\ \isakeyword{and}\ "valid\_path\ \isasymgamma "\ \isanewline
		\ \ \ \ \ \ \isakeyword{and}\ "z\isactrlsub 0\ \isasymnotin \ path\_image\ \isasymgamma "\ \isakeyword{and}\ "\isasymgamma \ 0\ =\ \isasymgamma \ 1"\isanewline
		\ \ \isakeyword{shows}\ "winding\_number\ \isasymgamma \ z\isactrlsub 0\ =\ -\ cindex\_pathE\ \isasymgamma \ z\isactrlsub 0\ /\ 2"
	\end{isabellebody}
\end{theorem}
\begin{proof}
	By assumption, we know that $\gamma$ is a loop, and the point $\gamma(0) = \gamma(1)$ can be away from the line $\{z \mid \Re(z) = \Re(z_0)  \}$ which makes Lemma \ref{thm:winding_number_cindex_pathE_aux} inapplicable. To resolve this problem, we look for a point $\gamma(s)$ on $\gamma$ such that $0 \leq s \leq 1$ and $\Re(\gamma(s)) = \Re(z_0)$, and we can either fail or succeed.

	In the case of failure, without loss of generality, we can assume $\Re(\gamma(t)) > \Re(z_0)$ for all $0 \leq t \leq 1$, and the shape of $\gamma$ is as illustrated in Fig.~\ref{fig:winding_cindex_path}a. As the path $\gamma$ does not cross the line $\{z \mid \Re(z) = \Re(z_0)  \}$, we can evaluate
	\[
	\Indp(\gamma,z_0) = 0
	\]
	\begin{equation*} \label{eq:winding_cindex_path1}
			n(\gamma,z_0)=\Re(n(\gamma,z_0)) = \frac{\Im(\Ln(\gamma(1) - z_0)) - \Im(\Ln(\gamma(0) - z_0))}{2 \pi} = 0
	\end{equation*}
	where $\Ln$ is the principle value of a complex logarithm function with its branch being the negative real axis and $-\pi<\Im(Ln(z)) \leq \pi$ for all $z$. Hence, $n(\gamma,z_0) = - \Indp(\gamma,z_0) / 2$ which concludes the case. 

	In the case of success, as illustrated in Fig. \ref{fig:winding_cindex_path}b, we have $\Re(\gamma(s)) = \Re(z_0)$.  We then define a shifted path $\gamma_s$:
	\[
	\gamma_s(t) =
	\begin{cases}
	\gamma(t+s) & \mbox{if } s+t \leq 1,\\
	\gamma(t+s-1) &  \mbox{ otherwise,}\\
	\end{cases}
	\]
	such that $\Re(\gamma_s(0)) = \Re(\gamma_s(1)) = \Re(z_0)$. By applying Lemma \ref{thm:winding_number_cindex_pathE_aux}, we obtain a relationship between $\Re(n(\gamma_s,z_0))$ and $\Indp(\gamma_s, z_0)$:
	\[
	2 \Re(n(\gamma_s,z_0)) =  - \Indp(\gamma_s,z_0),
	\]
	following which we have $n(\gamma,z_0) = - \Indp(\gamma,z_0) / 2$, since $n(\gamma_s,z_0) = n(\gamma,z_0)$ and $\Indp(\gamma_s,z_0) = \Indp(\gamma,z_0)$.
\end{proof}

\subsection{A Tactic for Evaluating Winding Numbers} \label{sec:tactic_for_evaluation}

With Proposition \ref{prop:winding_eq} formalised, we are now able to build a tactic to evaluate winding numbers using Cauchy indices. The idea has already been sketched in Examples \ref{ex:n_1_alt} and \ref{ex:n_0_alt}. We have built a tactic \isa{eval\_winding},  for goals of the form
\begin{equation} \label{eq:gamma_n_k}
	n(\gamma_1+\gamma_2+\cdots+\gamma_n,z_0) = k,
\end{equation}
where $k$ is an integer and $\gamma_j$ ($1 \leq j \leq n$) is either a linear path:
\[
\gamma_j(t) =  (1-t) a + t b \quad \mbox{ where } a, b \in \mathbb{C}
\]
or a part of a circular path:
\[
\gamma_j(t) = z+ r e^{i((1-t) a + t b)} \quad \mbox{ where } a, b, r \in \mathbb{R} \mbox{ and } z \in \mathbb{C}.
\]
The tactic \isa{eval\_winding} will transform (\ref{eq:gamma_n_k}) into 
\begin{equation} \label{itm:tac_1}
	\gamma_j(1) = \gamma_{j+1}(0) \mbox{ for all } 1 \leq j \leq n-1 \mbox{, and } \gamma_n(1) = \gamma_1(0),
\end{equation}
\begin{equation} \label{itm:tac_2} 
	 z_0 \not\in \{\gamma_j(t) \mid 0 \leq t \leq 1 \} \mbox{ for all } 1 \leq j \leq n,
\end{equation}
\begin{equation} \label{itm:tac_3} 
	\Indp(\gamma_1,z_0) + \Indp(\gamma_2,z_0) + \cdots + \Indp(\gamma_n,z_0) = -2 k,
\end{equation}
where (\ref{itm:tac_1}) ensures that the path $\gamma_1+\gamma_2+\cdots+\gamma_n$ is a loop; (\ref{itm:tac_2}) certifies that $z_0$ is not on the image of $\gamma_1+\gamma_2+\cdots+\gamma_n$.

To achieve this transformation, \isa{eval\_winding} will first perform a substitution step on the left-hand side of Equation (\ref{eq:gamma_n_k}) using Theorem \ref{thm:winding_number_cindex_pathE}. As the substitution is conditional, we will need to resolve four extra subgoals (i.e., (\ref{itm:tac_a}), (\ref{itm:tac_b}), (\ref{itm:tac_c}) and (\ref{itm:tac_d}) as follows) and Equation (\ref{eq:gamma_n_k}) is transformed into (\ref{itm:tac_e}):
\begin{equation} \label{itm:tac_a} 
	\mbox{
	 \isa{finite{\isacharunderscore}ReZ{\isacharunderscore}segments\ {\isacharparenleft}{\isasymgamma}\isactrlsub {\isadigit{1}}\ {\isacharplus}{\isacharplus}{\isacharplus}\ {\isasymgamma}\isactrlsub {\isadigit{2}}\ {\isacharplus}{\isacharplus}{\isacharplus}\ \isachardot\isachardot\isachardot\ \isacharplus\isacharplus\isacharplus {\isasymgamma}\isactrlsub n{\isacharparenright}\ z\isactrlsub {\isadigit{0}}},
	}
\end{equation}
\begin{equation}
	\label{itm:tac_b}
	\mbox{
	 \isa{valid{\isacharunderscore}path\ {\isacharparenleft}{\isasymgamma}\isactrlsub {\isadigit{1}}\ {\isacharplus}{\isacharplus}{\isacharplus}\ {\isasymgamma}\isactrlsub {\isadigit{2}}\ {\isacharplus}{\isacharplus}{\isacharplus}\ \isachardot\isachardot\isachardot\ \isacharplus\isacharplus\isacharplus {\isasymgamma}\isactrlsub n{\isacharparenright}},
	}
\end{equation}
\begin{equation}
	 \label{itm:tac_c} 
	 \mbox{\isa{z\isactrlsub {\isadigit{0}}\ {\isasymnotin}\ path{\isacharunderscore}image\ {\isacharparenleft}{\isasymgamma}\isactrlsub {\isadigit{1}}\ {\isacharplus}{\isacharplus}{\isacharplus}\ {\isasymgamma}\isactrlsub {\isadigit{2}}\ {\isacharplus}{\isacharplus}{\isacharplus}\ \isachardot\isachardot\isachardot\ \isacharplus\isacharplus\isacharplus {\isasymgamma}\isactrlsub n{\isacharparenright}},}
\end{equation}
\begin{equation}
	 \label{itm:tac_d} 
	 \mbox{\isa{{\isacharparenleft}{\isasymgamma}\isactrlsub {\isadigit{1}}\ {\isacharplus}{\isacharplus}{\isacharplus}\ {\isasymgamma}\isactrlsub {\isadigit{2}}\ {\isacharplus}{\isacharplus}{\isacharplus}\ \isachardot\isachardot\isachardot\ \isacharplus\isacharplus\isacharplus {\isasymgamma}\isactrlsub n{\isacharparenright}\ {\isadigit{0}}\ {\isacharequal}\ {\isacharparenleft}{\isasymgamma}\isactrlsub {\isadigit{1}}\ {\isacharplus}{\isacharplus}{\isacharplus}\ {\isasymgamma}\isactrlsub {\isadigit{2}}\ {\isacharplus}{\isacharplus}{\isacharplus}\ \isachardot\isachardot\isachardot\ \isacharplus\isacharplus\isacharplus {\isasymgamma}\isactrlsub n{\isacharparenright}\ {\isadigit{1}}},}
\end{equation}
\begin{equation}
	\label{itm:tac_e} 
	\mbox{\isa{{\isacharminus}\ cindex{\isacharunderscore}pathE\ {\isacharparenleft}{\isasymgamma}\isactrlsub {\isadigit{1}}\ {\isacharplus}{\isacharplus}{\isacharplus}\ {\isasymgamma}\isactrlsub {\isadigit{2}}\ {\isacharplus}{\isacharplus}{\isacharplus}\ \isachardot\isachardot\isachardot\ \isacharplus\isacharplus\isacharplus {\isasymgamma}\isactrlsub n{\isacharparenright}\ z\isactrlsub {\isadigit{0}}\ {\isacharslash}\ {\isadigit{2}}\ {\isacharequal}\ k}.}
\end{equation}
To simplify (\ref{itm:tac_a}), the tactic will keep applying the following introduction rule:\footnote{Applying an introduction rule will replace a goal by a set of subgoals derived from the premises of the rule, provided the goal can be unified with the conclusion of the rule.}
\begin{lemma}[\isa{finite\_ReZ\_segments\_joinpaths}]
	\label{thm:finite_ReZ_segments_joinpaths}
	\vspace{-7pt}
	\begin{isabellebody}
		\isanewline
		\ \ \isakeyword{fixes}\ \isasymgamma \isactrlsub 1\ \isasymgamma \isactrlsub 2\ ::\ "real\ \isasymRightarrow \ complex"\ \isakeyword{and}\ z\isactrlsub 0\ ::\ complex\isanewline
		\ \ \isakeyword{assumes}\ "finite\_ReZ\_segments\ \isasymgamma \isactrlsub 1\ z\isactrlsub 0"\ \isakeyword{and}\ "finite\_ReZ\_segments\ \isasymgamma \isactrlsub 2\ z\isactrlsub 0"\ \isanewline
		\ \ \ \ \isakeyword{and}\ "path\ \isasymgamma \isactrlsub 1"\ \isakeyword{and}\ "path\ \isasymgamma \isactrlsub 2"\ \isakeyword{and}\ "\isasymgamma \isactrlsub 1\ 1\ =\ \isasymgamma \isactrlsub 2\ 0"\isanewline
		\ \ \isakeyword{shows}\ "finite\_ReZ\_segments\ (\isasymgamma \isactrlsub 1+++\isasymgamma \isactrlsub 2)\ z\isactrlsub 0"
	\end{isabellebody}
\end{lemma}
\noindent to eliminate the path join operations (\isa{\isacharplus\isacharplus\isacharplus}) until the predicate \isa{finite\_ReZ\_segments} is only applied to a linear path or a part of a circular path, and either of these two cases can be directly discharged because these two kinds of paths are proved to be divisible into a finite number of segments by the imaginary axis:
\begin{lemma} [\isa{finite\_ReZ\_segments\_linepath}]
	\vspace{-7pt}
	\begin{isabellebody}
		\isanewline
		\ \ "finite\_ReZ\_segments\ (linepath\ a\ b)\ z"
	\end{isabellebody}
\end{lemma}
\begin{lemma} [\isa{finite\_ReZ\_segments\_part\_circlepath}]
	\vspace{-7pt}
	\begin{isabellebody}
		\isanewline
		\ \ "finite\_ReZ\_segments\ (part\_circlepath\ z0\ r\ st\ tt)\ z"
	\end{isabellebody}
\end{lemma}
\noindent In terms of other subgoals introduced when applying Lemma \ref{thm:finite_ReZ_segments_joinpaths}, such as \isa{path\ \isasymgamma \isactrlsub 1}, \isa{path\ \isasymgamma \isactrlsub 2} and  \isa{\isasymgamma \isactrlsub 1\ 1\ =\ \isasymgamma \isactrlsub 2\ 0}, we can discharge them by the following introduction and simplification rules (all of which have been formally proved):
\begin{itemize}
	\item \isa{\isasymlbrakk path\ \isasymgamma \isactrlsub 1;\ path\ \isasymgamma \isactrlsub 2;\ \isasymgamma \isactrlsub 1\ 1\ =\ \isasymgamma \isactrlsub 2\ 0\isasymrbrakk \ \isasymLongrightarrow \ path(\isasymgamma \isactrlsub 1\ +++\ \isasymgamma \isactrlsub 2)},
	\item \isa{path\ (part\_circlepath\ z\isactrlsub0\ r\ st\ tt)},
	\item \isa{path\ (linepath\ a\ b)},
	\item \isa{(\isasymgamma \isactrlsub 1\ +++\ \isasymgamma \isactrlsub 2)\ 1\ =\ \isasymgamma \isactrlsub 2\ 1},
	\item \isa{(\isasymgamma \isactrlsub 1\ +++\ \isasymgamma \isactrlsub 2)\ 0\ =\ \isasymgamma \isactrlsub 1\ 0}.
\end{itemize}
As a result, \isa{eval\_winding} will eventually simplify the subgoal (\ref{itm:tac_a}) to (\ref{itm:tac_1}).

Similar to the process of simplifying (\ref{itm:tac_a}) to (\ref{itm:tac_1}), the tactic \isa{eval\_winding} will
also simplify
\begin{itemize}
	\item (\ref{itm:tac_b}) to  (\ref{itm:tac_1}),
	\item (\ref{itm:tac_c}) to  (\ref{itm:tac_2}),
	\item and (\ref{itm:tac_d}) to (\ref{itm:tac_1}).
\end{itemize}
Finally, with respect to (\ref{itm:tac_e}), we can similarly rewrite with a rule between the Cauchy index (\isa{cindex\_pathE}) and the path join operation (\isa{\isacharplus\isacharplus\isacharplus}):
\begin{lemma} [\isa{cindex\_pathE\_joinpaths}] \label{thm:cindex_pathE_joinpaths}
	\vspace{-7pt}
	\begin{isabellebody}
		\isanewline
		\ \ \isakeyword{fixes}\ \isasymgamma \isactrlsub 1\ \isasymgamma \isactrlsub 2\ ::\ "real\ \isasymRightarrow \ complex"\ \isakeyword{and}\ z\isactrlsub 0\ ::\ complex\isanewline
		\ \ \isakeyword{assumes}\ "finite\_ReZ\_segments\ \isasymgamma \isactrlsub 1\ z\isactrlsub 0"\ \isakeyword{and}\ "finite\_ReZ\_segments\ \isasymgamma \isactrlsub 2\ z\isactrlsub 0"\ \isanewline
		\ \ \ \ \isakeyword{and}\ "path\ \isasymgamma \isactrlsub 1"\ \isakeyword{and}\ "path\ \isasymgamma \isactrlsub 2"\ \isakeyword{and}\ "\isasymgamma \isactrlsub 1\ 1\ =\ \isasymgamma \isactrlsub 2\ 0"\isanewline
		\ \ \isakeyword{shows}\ "cindex\_pathE\ (\isasymgamma \isactrlsub 1\ +++\ \isasymgamma \isactrlsub 2)\ z\isactrlsub 0\ =\ cindex\_pathE\ \isasymgamma \isactrlsub 1\ z\isactrlsub 0\ +\ cindex\_pathE\ \isasymgamma \isactrlsub 2\ z\isactrlsub 0"
	\end{isabellebody}
\end{lemma}
\noindent to convert the subgoal (\ref{itm:tac_e}) to (\ref{itm:tac_1}) and (\ref{itm:tac_3}).

After building the tactic \isa{eval\_winding}, we are now able to convert a goal like Equation (\ref{eq:gamma_n_k}) to (\ref{itm:tac_1}), (\ref{itm:tac_2}) and (\ref{itm:tac_3}). In most cases, discharging (\ref{itm:tac_1}) and (\ref{itm:tac_2}) is straightforward. To derive (\ref{itm:tac_3}), we will need to formally evaluate each $\Indp(\gamma_j,z_0)$ ($1 \leq j \leq n$) when $\gamma_j$ is either a linear path or a part of a circular path.

When $\gamma_j$ is a linear path, the following lemma grants us a way to evaluate $\Indp(\gamma_j,z_0)$ through its right-hand side:
\begin{lemma} [\isa{cindex\_pathE\_linepath}] \label{thm:cindex_pathE_linepath}
	\vspace{-7pt}
	\begin{isabellebody}
		\isanewline
		\ \ \isakeyword{fixes}\ a\ b\ z\isactrlsub 0\ ::\ complex\isanewline
		\ \ \isakeyword{assumes}\ "z\isactrlsub 0\isasymnotin path\_image\ (linepath\ a\ b)"\isanewline
		\ \ \isakeyword{shows}\ "cindex\_pathE\ (linepath\ a\ b)\ z\isactrlsub 0\ =\ (\isanewline
		\ \ \ \ let\ c1\ =\ Re\ a\ -\ Re\ z\isactrlsub 0;\ \isanewline
		\ \ \ \ \ \ \ \ c2\ =\ Re\ b\ -\ Re\ z\isactrlsub 0;\ \isanewline
		\ \ \ \ \ \ \ \ c3\ =\ Im\ a\ *\ Re\ b\ +\ Re\ z\isactrlsub 0\ *\ Im\ b\ +\ Im\ z\isactrlsub 0\ *\ Re\ a\ -\ Im\ z\isactrlsub 0\ *\ Re\ b\isanewline
		\ \ \ \ \ \ \ \ \ \ \ \ \ \ -\ Im\ b\ *\ Re\ a\ -\ Re\ z\isactrlsub 0\ *\ Im\ a;\isanewline
		\ \ \ \ \ \ \ \ d1\ =\ Im\ a\ -\ Im\ z\isactrlsub 0;\isanewline
		\ \ \ \ \ \ \ \ d2\ =\ Im\ b\ -\ Im\ z\isactrlsub 0\isanewline
		\ \ \ \ in\ if\ (c1>0\ \isasymand \ c2<0)\ \isasymor \ (c1<0\ \isasymand \ c2>0)\ then\isanewline
		\ \ \ \ \ \ \ \ \ \ (if\ c3>0\ then\ 1\ else\ -1)\ \isanewline
		\ \ \ \ \ \ \ else\ \isanewline
		\ \ \ \ \ \ \ \ \ \ (if\ (c1=0\ \isasymlongleftrightarrow \ c2\isasymnoteq 0)\ \isasymand \ (c1=0\ \isasymlongrightarrow d1\isasymnoteq 0)\ \isasymand \ (c2=0\ \isasymlongrightarrow \ d2\isasymnoteq 0)\ then\isanewline
		\ \ \ \ \ \ \ \ \ \ \ \ \ \ if\ (c1=0\ \isasymand \ (c2\ >0\ \isasymlongleftrightarrow \ d1>0))\ \isasymor \ (c2=0\ \isasymand \ (c1\ >0\ \isasymlongleftrightarrow \ d2<0))\isanewline
		\ \ \ \ \ \ \ \ \ \ \ \ \ \ then\ 1/2\ else\ -1/2\isanewline
		\ \ \ \ \ \ \ \ \ \ else\ 0))"
	\end{isabellebody}
\end{lemma}
Although Lemma \ref{thm:cindex_pathE_linepath} may appear terrifying, evaluating its right-hand side is usually automatic when the number of free variables is small. For example, in a formal proof of Example \ref{ex:n_1_alt} in Isabelle/HOL, we can have the following fragment:

\begin{isabelle}
	\isacommand{lemma}	\ \isanewline
	\ \ \isakeyword{fixes}\ R::real\isanewline
	\ \ \isakeyword{assumes}\ "R>1"\isanewline
	\ \ \isakeyword{shows}\ "winding\_number\ (part\_circlepath\ 0\ R\ 0\ pi\ +++\ linepath\ (-R)\ R)\ \isasymi \ =\ 1"\isanewline
	\isacommand{proof}	\ (winding\_eval, simp\_all)\isanewline
	\ \ \isachardot\isachardot\isachardot \isanewline
	\ \ \isacommand{have}\ "\isasymi \ \isasymnotin \ path\_image\ (linepath\ (-\ R)\ (R::complex))"\ \isacommand{by} \isachardot\isachardot\isachardot 	\isanewline
	\ \ \isacommand{from}\ cindex\_pathE\_linepath[OF\ this]\ \isacartoucheopen R>1\isacartoucheclose \isanewline
	\ \ \isacommand{have}\ "cindex\_pathE\ (linepath\ (-R)\ (R::complex))\ \isasymi \ =\ -1"\ \isacommand{by}	\ auto\isanewline
	\ \ \isachardot\isachardot\isachardot \isanewline
	\isacommand{qed}
\end{isabelle}
where \isa{winding\_eval} is first applied to convert the goal into (\ref{itm:tac_1}), (\ref{itm:tac_2}) and (\ref{itm:tac_3}), and \isa{simp\_all} subsequently simplifies those newly generated subgoals. In the middle of the proof, we show that the complex point $i$ is not on the image of the linear path $L_r$ (i.e., \isa{linepath\ (-R)\ (R::complex))} in Isabelle/HOL), following which we apply Lemma \ref{thm:cindex_pathE_linepath} to derive $\Indp(L_r,i) = -1$: the evaluation process is automatic through the command \isa{auto}, given the assumption \isa{R>1}.

When $\gamma_j$ is a part of a circular path, a similar lemma has been provided to facilitate the evaluation of $\Indp(\gamma_j,z_0)$.
\subsection{Subtleties} \label{sec:subtleties}

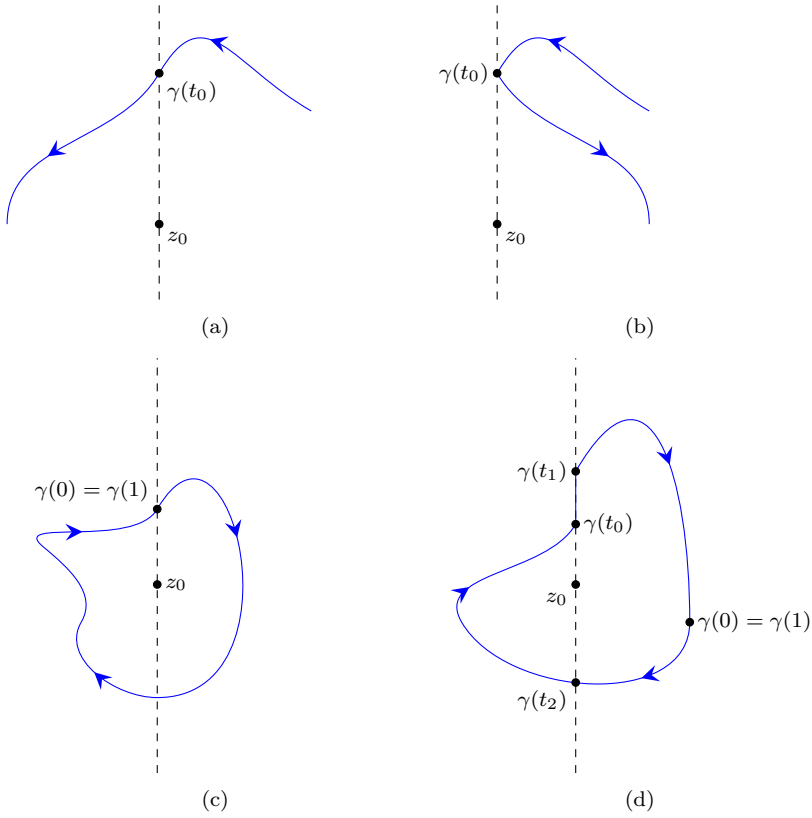
\begin{figure}[ht]
	\centering
	\begin{subfigure}[t]{0.45\textwidth}
		\begin{tikzpicture}[scale=1]

		\draw [blue] (2,1.5) to[out=150,in=60,looseness=1.5] (0,2) to[out=240,in=90,looseness=1] (-2,0)
		[arrow inside={end=stealth,opt={,scale=2}}{0.3,0.8}];

		\draw[dashed] (0,-1)--(0,3) node[right]{};

		\draw [fill] (0,2) circle [radius=.5mm] node [below right] {$\gamma(t_0)$};

		\draw [fill] (0,0) circle [radius=.5mm] node [below right] {$z_0$};

		\end{tikzpicture}
		\caption{}
	\end{subfigure}
	\begin{subfigure}[t]{0.45\textwidth}
		\begin{tikzpicture}[scale=1]

		\draw [blue] (2,1.5) to[out=150,in=60,looseness=1.5] (0,2) to[out=300,in=90,looseness=1] (2,0)
		[arrow inside={end=stealth,opt={,scale=2}}{0.3,0.8}];

		\draw[dashed] (0,-1)--(0,3) node[right]{};

		\draw [fill] (0,2) circle [radius=.5mm] node [left] {$\gamma(t_0)$};

		\draw [fill] (0,0) circle [radius=.5mm] node [below right] {$z_0$};

		\end{tikzpicture}
		\caption{}
	\end{subfigure}
	\begin{subfigure}[t]{0.45\textwidth}
		\begin{tikzpicture}[scale=1]
		\draw[dashed] (0,-2)--(0,3.5) node[right]{};

		\draw [blue] (0,-1) to[out=180,in=240,looseness=1] (-1,0) to[out=60,in=320,looseness=1] (-1.5,1)
		to[out=140,in=240,looseness=1] (0,1.5) to[out=60,in=0,looseness=2] (0,-1) [arrow inside={end=stealth,opt={,scale=2}}{.1,.4,.7}];


		\draw [fill] (0,1.5) circle [radius=.5mm] node [above left] {$\gamma(0) = \gamma(1)$};

		\draw [fill] (0,.5) circle [radius=.5mm] node [right] {$z_0$};

		\end{tikzpicture}
		\caption{}
	\end{subfigure}
	\begin{subfigure}[t]{0.45\textwidth}
		\begin{tikzpicture}[scale=1]
		\draw[dashed] (0,-2)--(0,3.5) node[right]{};

		\draw [blue] (1.5,0) to[out=270,in=300,looseness=1] (-1.5,0) to[out=120,in=240,looseness=1]  (0,1.3) -- (0,2) to[out=60,in=90,looseness=2] (1.5,0) [arrow inside={end=stealth,opt={,scale=2}}{.1,.4,.8}];

		\draw [fill] (0,2) circle [radius=.5mm] node [left] {$\gamma(t_1)$} ;

		\draw [fill] (0,1.3) circle [radius=.5mm] node [right] {$\gamma(t_0)$};

		\draw [fill] (0,-.8) circle [radius=.5mm] node [below left] {$\gamma(t_2)$};

		\draw [fill] (0,.5) circle [radius=.5mm] node [below left] {$z_0$};

		\draw [fill] (1.5,0) circle [radius=.5mm] node [right] {$\gamma(0) = \gamma(1)$};

		\end{tikzpicture}
		\caption{}
	\end{subfigure}
	\caption{Different ways a path $\gamma$ can intersect with the line $\{z \mid \Re(z) = \Re(z_0)\}$}
	\label{fig:subtleties}
\end{figure}

The first subtlety we have encountered during the formalisation of Proposition \ref{prop:winding_eq} is about the definitions of jumps and Cauchy indices, for which our first attempt followed the standard definitions in textbooks \cite{Basu:2006bo,Marden:1949ui,Rahman:2016us}.

\begin{definition} [Jump] \label{def:old_jump}
	For $f : \mathbb{R} \rightarrow \mathbb{R}$ and $x \in \mathbb{R}$, we define
	\[
	\jump(f,x) =
	\begin{cases}
	1 & \mbox{if } \lim_{u \rightarrow x^-} f(u)=- \infty \mbox{ and } \lim_{u \rightarrow x^+} f(u)=+\infty,\\
	-1& \mbox{if } \lim_{u \rightarrow x^-} f(u)=+\infty \mbox{ and } \lim_{u \rightarrow x^+} f(u)=-\infty,\\
	0 &  \mbox{ otherwise.}\\
	\end{cases}
	\]
\end{definition}

\begin{definition}[Cauchy index] \label{def:old_ind}
	For $f : \mathbb{R} \rightarrow \mathbb{R}$ and $a, b \in \mathbb{R}$, the Cauchy index of $f$ over an open interval $(a,b)$ is defined as
	\[
	\Ind_a^b(f) = \sum_{x \in (a,b)} \jump(f,x).
	\]
\end{definition}

The impact of the difference between the current definition of the Cauchy index (i.e., Definition~\ref{def:ind}) and the classic one (i.e., Definition~\ref{def:old_ind}) is small when formalising the Sturm-Tarski theorem \cite{Li:2017cg,Sturm_Tarski-AFP}, where $f$ is a rational function. In this case, the path $\gamma$ intersects with the line $\{z \mid \Re(z) = \Re(z_0)\}$ a finite number of times, and for each intersection point (see Fig.~\ref{fig:subtleties}a and b), by letting $f(t) = {\Im(\gamma(t) - z_0)}/{\Re(\gamma(t) - z_0)}$, we have
\[
\jump(f,t) = \jump_+(f,t) - \jump_-(f,t),
\]
hence
\[
\sum_{x \in (a,b)} \jump(f,x) = \sum_{x \in [a,b)} \jump_+(f,x) - \sum_{x \in (a,b]} \jump_-(f,x),
\]
provided $\jump_+(f,a) =0$ and $\jump_-(f,b) = 0$. That is, the classic Cauchy index  and the current one are equal when $f$ is a rational function and does not jump at both ends of the target interval.

Naturally, the disadvantages of Definition \ref{def:old_ind} are twofold:
\begin{itemize}
	\item The function $\lambda t.\, \Re(\gamma(t) - z_0)$ cannot vanish at either end of the interval. That is, we need to additionally assume $\Re(\gamma(0) - z_0) \neq 0$ as in Rahman and Schmeisser's formulation \cite[Lemma 11.1.1 and Theorem 11.1.3]{Rahman:2016us}, and Proposition \ref{prop:winding_eq} will be inapplicable in the case of Fig.~\ref{fig:subtleties}c where $\Re(\gamma(0)) =  \Re(\gamma(1)) = \Re(z_0)$.
	\item The function $\lambda t.\, {\Im(\gamma(t) - z_0)}/{\Re(\gamma(t) - z_0)}$ has to be rational, which makes Proposition~\ref{prop:winding_eq} inapplicable for cases like in Fig.~\ref{fig:subtleties}d (if we follow Definition~\ref{def:old_ind}). To elaborate, it can be observed  in Fig.~\ref{fig:subtleties}d that $n(\gamma,z_0) = -1$, while we will only get a wrong answer by following Definition~\ref{def:old_ind} and evaluating via Proposition~\ref{prop:winding_eq}:
	\[
	- \frac{1}{2} \left( \sum_{x \in (0,1)} \jump(f,x) \right) = - \frac{\jump(f,t_2)}{2} = - \frac{1}{2},
	\]
	where $f(t) = {\Im(\gamma(t) - z_0)}/{\Re(\gamma(t) - z_0)}$.
	In comparison, Definition~\ref{def:ind} leads to the correct answer:
	\[
	\begin{split}
	n(\gamma,z_0) &= - \frac{1}{2} \left(  \sum_{x \in [0,1)} \jump_+(f,x) - \sum_{x \in (0,1]} \jump_-(f,x)  \right)\\
	& = - \frac{1}{2} \left( \jump_+(f,t_2) + \jump_+(f,t_1) - \jump_-(f,t_2) - \jump_-(f,t_0) \right) \\
	& = - \frac{1}{2} \left( \frac{1}{2} + \frac{1}{2} - (- \frac{1}{2}) - (- \frac{1}{2}) \right) \\
	& = - 1.
	\end{split}
	\]
\end{itemize}
Fortunately, Michael Eisermann \cite{MichaelEisermann:2012be} recently proposed a new formulation of the Cauchy index that overcomes those two disadvantages, and this new formulation is what we have followed (in Definitions \ref{def:jumpF} and \ref{def:ind}).

Another subtlety we ran into was the well-definedness of the Cauchy index. Such well-definedness is usually not an issue and left implicit in the literature, because, in most cases, the Cauchy index is only defined on rational functions, where only finitely many points can contribute to the sum. When attempting to formally derive Proposition \ref{prop:winding_eq}, we realised that this assumption needed to be made explicit, since the path $\gamma$ can be flexible enough to allow the function $f(t) = {\Im(\gamma(t) - z_0)}/{\Re(\gamma(t) - z_0)}$ to be non-rational (e.g. Fig.~\ref{fig:subtleties}d). In our first attempt of following Definition \ref{def:old_ind}, the Cauchy index was formally defined as follows:
\begin{isabelle}
	\isacommand{definition}	\ cindex::"real\ \isasymRightarrow \ real\ \isasymRightarrow \ (real\ \isasymRightarrow \ real)\ \isasymRightarrow \ int"\ \isakeyword{where}\isanewline
	\ \ "cindex\ a\ b\ f\ =\ (\isasymSum x\isasymin \isacharbraceleft x.\ jump\ f\ x\isasymnoteq 0\ \isasymand \ a<x\ \isasymand \ x<b\isacharbraceright .\ jump\ f\ x)"
\end{isabelle}
and its well-definedness was ensured by the finite number of times that $\gamma$ crosses the line $\{z \mid \Re(z) = \Re(z_0) \}$:
\begin{isabelle}
	\isacommand{definition}	\ finite\_axes\_cross::"(real\ \isasymRightarrow \ complex)\ \isasymRightarrow \ complex\ \isasymRightarrow \ bool"\ \isakeyword{where}\isanewline
	\ \ "finite\_axes\_cross\ \isasymgamma \ z\isactrlsub 0\ =\ \isanewline
	\ \ \ \ \ finite\ \isacharbraceleft t.\ (Re\ (\isasymgamma \ t\ -\ z\isactrlsub 0)\ =\ 0\ \isasymor \ Im\ (\isasymgamma \ t\ -\ z\isactrlsub 0)\ =\ 0)\ \isasymand \ 0\ \isasymle \ t\ \isasymand \ t\ \isasymle \ 1\isacharbraceright "
\end{isabelle}
where the part \isa{Re\ (\isasymgamma \ t\ -\ z\isactrlsub 0)\ =\ 0} ensures that \isa{jump f t} is non-zero only at finitely many points over the interval $[0,1]$. When constrained by \isa{finite\_axes\_cross}, the function $f(t) = {\Im(\gamma(t) - z_0)}/{\Re(\gamma(t) - z_0)}$ behaves like a rational function. More importantly, the path $\gamma$, in this case, can be divided into a finite number of ordered segments delimited by those points over $[0,1]$, which makes an inductive proof of Proposition \ref{prop:winding_eq} possible. However, after abandoning our first attempt and switching to Definition \ref{def:ind}, the well-definedness of the Cauchy index is assured by the finite number of $\jump_+$ and $\jump_-$ of $f$ (i.e., Definition \isa{finite\_jumpFs} in  \S\ref{sec:formal_proof_prop}), with which we did not know how to divide the path $\gamma$ into segments and carry out an inductive proof. It took us some time to properly define the assumption of a finite number of segments (i.e., Definition \isa{finite\_ReZ\_segments}) that implied the well-definedness using Lemma \ref{thm:finite_ReZ_segments_imp_jumpFs} and provided a lemma for inductive proofs (i.e., Lemma \ref{thm:finite_ReZ_segments_induct}).

\section{Counting the Number of Complex Roots} \label{sec:root_counting}

The previous section described a way to evaluate winding numbers via Cauchy indices. In this section, we will further explore this idea and propose verified procedures to count the number of complex roots of a polynomial in some domain such as a rectangle and a half-plane.

Does a winding number have anything to do with the number of roots of a polynomial? The answer is yes. Thanks to the argument principle, we can calculate the number of roots by evaluating a contour integral:
\begin{equation} \label{eq:count_roots_intro1}
\frac{1}{2 \pi i} \oint_\gamma \frac{p'(x)}{p(x) } d x = N
\end{equation}
where $p \in \mathbb{C}[x]$, $p'(x)$ is the first derivative of $p$ and $N$ is the number of complex roots of $p$ (counted with multiplicity) inside the loop $\gamma$. Also, by the definition of winding numbers, we have
\begin{equation}  \label{eq:count_roots_intro2}
n(p \circ \gamma,0) = 	\frac{1}{2 \pi i} \oint_\gamma \frac{p'(x)}{p(x) } d x.
\end{equation}
Combining Equations (\ref{eq:count_roots_intro1}) and  (\ref{eq:count_roots_intro2}) gives us the relationship between a winding number and the number of roots of a polynomial:
\begin{equation} \label{eq:count_roots_intro3}
n(p \circ \gamma,0)  = N.
\end{equation}
And the question becomes: can we evaluate $n(p \circ \gamma,0)$ via Cauchy indices?

\subsection{Roots in a Rectangle}
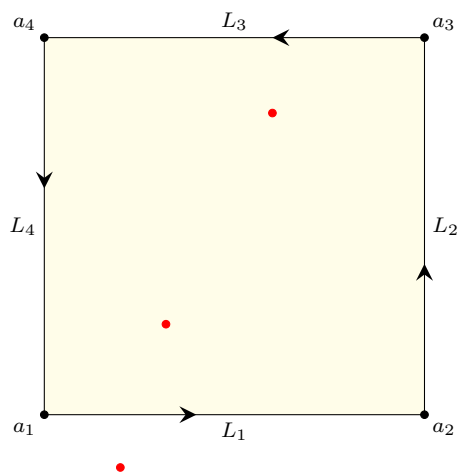
\begin{figure}[ht]
	\centering
	\begin{tikzpicture}[scale=1]
	\fill [yellow!10] (0,0) rectangle (5,5);
	\draw (0,0) -- (5,0) -- (5,5) -- (0,5) -- (0,0)
	[arrow inside={end=stealth,opt={,scale=2}}{0.1,.35,.6,0.85}];

	\draw [fill,red] (-2,4) circle [radius=.5mm] node (r1) {};
	\draw [fill,red] (1.6,1.2) circle [radius=.5mm] node (r2) {} ;
	\draw [fill,red] (1,-.7) circle [radius=.5mm] node (r3) {};
	\draw [fill,red] (3,4) circle [radius=.5mm] node (r4) {};

	\draw [fill] (0,0) circle [radius=.5mm] node [below left] {$a_1$};
	\draw [fill] (5,0) circle [radius=.5mm] node [below right] {$a_2$};
	\draw [fill] (5,5) circle [radius=.5mm] node [above right] {$a_3$};
	\draw [fill] (0,5) circle [radius=.5mm] node [above left] {$a_4$};

	%

	\node at (2.5,0) [below] {$L_1$};
	\node at (5,2.5) [right] {$L_2$};
	\node at (2.5,5) [above] {$L_3$};
	\node at (0,2.5) [left] {$L_4$};
	\end{tikzpicture}
	\caption{Complex roots of a polynomial (red dots) and a rectangular path ($L_1+L_2+L_3+L_4$) on the complex plane}
	\label{fig:roots_rectangle}
\end{figure}
Let $N$ be the number of complex roots of a polynomial $p$ inside the rectangle defined by its lower left corner $a_1$ and upper right corner $a_3$. As illustrated in Fig. \ref{fig:roots_rectangle}, we can define four linear paths along the edge of the rectangle:
\[
\begin{split}
L_1(t) &=(1-t) a_1 + t a_2\\
L_2(t) &=(1-t) a_2 + t a_3\\
L_3(t) &=(1-t) a_3 + t a_4\\
L_4(t) &=(1-t) a_4 + t a_1\\
\end{split}
\]
where $a_2 = \Re(a_3) + i \Im(a_1)$ and $a_4 = \Re(a_1) + i \Im(a_3)$. Combining Proposition \ref{prop:winding_eq} with Equation (\ref{eq:count_roots_intro3}) yields
\begin{equation} \label{eq:roots_rectangle1}
\begin{split}
N & = n(p \circ (L_1+L_2+L_3+L_4),0) \\
& =  -\frac{1}{2} \Indp(p \circ (L_1+L_2+L_3+L_4),0) \\
& =  -\frac{1}{2} \left(\Indp(p \circ L_1,0)+ \Indp(p \circ L_2,0)+ \Indp(p \circ L_3,0) + \Indp(p \circ L_4,0)\right).\\
\end{split}
\end{equation}
Here, the path $p \circ L_j : [0,1] \rightarrow \mathbb{C}$ ($1 \leq j \leq 4$) is (mostly) neither a linear path nor a part of a circular path, which indicates that the evaluation strategies of \S\ref{sec:tactic_for_evaluation}, such as Lemma \ref{thm:cindex_pathE_linepath}, will no longer apply. Thankfully, the Sturm-Tarski theorem \cite{Sturm_Tarski-AFP,Li:2017cg} came to our rescue.

In general, the Sturm-Tarski theorem is about calculating Tarski queries through sign variations and signed remainder sequences: let $p,q \in \mathbb{R}[x]$, $a$ and $b$ be two extended real numbers such that $a<b$ and are not roots of $p$, we have
\begin{equation} \label{eq:roots_rectangle1_2}
	 \TaQ(q,p,a,b)  = \Var(\SRemS(p,p' q);a,b)
\end{equation}
\noindent where 
\begin{itemize}
	\item $p'$ is the first derivative of $p$,
	\item the Tarski query $\TaQ(q,p,a,b)$ defined as follows:
		\[
			 \TaQ(q,p,a,b)  = \sum_{x \in (a,b), p(x) = 0} \sgn(q(x)),
		\]
	\item $\SRemS(p,q)$ is the signed remainder sequence started with $p$ and $q$.
	\item Let $[p_1,p_2,...,p_n]$ be a sequence of polynomials, $\Var([p_1,p_2,...,p_n];a,b)$ is the difference in the number of sign variations when evaluating $[p_1,p_2,...,p_n]$ at $a$ and $b$:
	\begin{multline}
		\Var([p_1,p_2,...,p_n];a,b) \\= 	\Var([p_1(a),p_2(a),...,p_n(a)]) - \Var([p_1(b),p_2(b),...,p_n(b)]).
	\end{multline}
\end{itemize}
Note that when $q=1$, (\ref{eq:roots_rectangle1_2}) becomes the famous Sturm's theorem, which counts the number of distinct real roots over an interval. For example, by calculating
\[
\begin{aligned}
\TaQ(1,(x-1)(x-2),0,3) & = \Var(\SRemS(x^2-3x+2,2 x - 3);0,3) \\
	& = \Var([x^2-3x+2,2 x - 3,1/4];0,3) \\
	& = \Var([x^2-3x+2,2 x - 3,1/4];0)  \\
	& \qquad - \Var([x^2-3x+2,2 x - 3,1/4];3) \\
	& = \Var([2,-3,1/4]) - \Var([2,3,1/4]) \\
	& = 2 - 0 = 2,
\end{aligned}
\]
we know that the polynomial $x^2-3x+2$ has two distinct real roots within the interval $(0,3)$.

In our previous formal proof of the Sturm-Tarski theorem \cite{Sturm_Tarski-AFP,Li:2017cg}, we used the Cauchy index to relate the Tarski query and the right-hand side of (\ref{eq:roots_rectangle1_2}). Therefore, as a byproduct, we can also evaluate the Cauchy index through sign variations and signed remainder sequences:
\begin{equation} \label{eq:roots_rectangle2}
\Ind_a^b \left(\lambda t.\, \frac{q(t)}{p(t)} \right) = \Var(\SRemS(p,q);a,b),
\end{equation}
\noindent where $p,q \in \mathbb{R}[x]$, $a, b$ are two extended real numbers such that $a<b$ and are not roots of $p$.

Back to the case of $\Indp(p \circ L_j,0)$, we have
\[
\Indp(p \circ L_j,0) = \Ind_0^1\left(\lambda t.\ \frac{\Im(p(L_j(t)))}{\Re(p(L_j(t)))} \right),
\]
and both $\Im(p(L_j(t)))$ and $\Re(p(L_j(t)))$ happen to be polynomials with real coefficients. Therefore, combining Equations (\ref{eq:roots_rectangle1}) and (\ref{eq:roots_rectangle2}) yields an approach to count the number of roots inside a rectangle.

While proceeding to the formal development, the first problem we encountered was that the Cauchy index in Equation (\ref{eq:roots_rectangle2}) actually follows the classic definition (i.e., Definition \ref{def:old_ind}), and is different from the one in Equation (\ref{eq:roots_rectangle1}) (i.e., Definitions \ref{def:ind} and \ref{def:indp}). Subtle differences between these two formulations have already been discussed in \S\ref{sec:subtleties}. Luckily, Eisermann \cite{MichaelEisermann:2012be} has also described an alternative sign variation operator so that our current definition of the Cauchy index (i.e., Definition \ref{def:ind}) can be computationally evaluated:
\begin{lemma} [\isa{cindex\_polyE\_changes\_alt\_itv\_mods}]
	\label{thm:cindex_polyE_changes_alt_itv_mods}
	\vspace{-7pt}
	\begin{isabellebody}
		\isanewline
		\ \ \isakeyword{fixes}\ a\ b::real\ \isakeyword{and}\ p\ q::"real\ poly"\isanewline
		\ \ \isakeyword{assumes}\ "a\ <\ b"\ \isakeyword{and}\ "coprime\ p\ q"\isanewline
		\ \ \isakeyword{shows}\ "cindex\_polyE\ a\ b\ q\ p\ =\ changes\_alt\_itv\_smods\ a\ b\ p\ q\ /\ 2"
	\end{isabellebody}
\end{lemma}
\noindent Here, \isa{cindex\_polyE a b q p} encodes our current definition of the Cauchy index $\Ind^b_a(\lambda t.\, q(t) / p(t))$, and \isa{changes\_alt\_itv\_smods\ a\ b\ p\ q} stands for
\begin{equation} \label{eq:roots_rectangle3}
	\widehat{\Var}(\SRemS(p,q);a,b)
\end{equation}
\noindent where the alternative sign variation operator $\widehat{\Var}$ is defined as follows:
\[
\begin{aligned} 
	\widehat{\Var}([p_1,p_2,...,p_3];a,b) &= \widehat{\Var}([p_1,p_2,...,p_3];a) - \widehat{\Var}([p_1,p_2,...,p_3];b), \\
	\widehat{\Var}([p_1,p_2,...,p_3];a) & = \widehat{\Var}([p_1(a),p_2(a),...,p_3(a)]), \\
	\widehat{\Var}([]) &= 0,\\
	\widehat{\Var}([x_1]) &= 0,\\
	\widehat{\Var}([x_1,x_2,...,x_n]) &= |\sgn(x_1) - \sgn(x_2)| + \widehat{\Var}([x_2,...,x_n]).\\
\end{aligned}
\]
The difference between $\widehat{\Var}$ and $\Var$ is that $\Var$ discards zeros before calculating variations while $\widehat{\Var}$ takes zeros into consideration. For example, $\Var([1,0,-2]) = \Var([1,-2]) = 1$, while $\widehat{\Var}([1,0,-2]) = 2$.

Before implementing Equation (\ref{eq:roots_rectangle1}), we need to realise that there is a restriction in our strategy: roots are not allowed on the border (i.e., the image of the path $L_1+L_2+L_3+L_4$). To computationally check this restriction, the following function is defined
\begin{isabelle}
	\isacommand{definition}	\ no\_proots\_line::"complex\ poly\ \isasymRightarrow \ complex\ \isasymRightarrow \ complex\ \isasymRightarrow \ bool"\ \isakeyword{where}\isanewline
	\ \ "no\_proots\_line\ p\ a\ b\ =\ (proots\_within\ p\ (closed\_segment\ a\ b)\ =\ \isacharbraceleft \isacharbraceright )"
\end{isabelle}
which will return ``true'' if there is no root on the closed segment between \isa{a} and \isa{b}, and ``false'' otherwise. Here, \isa{closed\_segment a b} is defined as the set $\{(1-u) a + u b \mid 0 \leq u \leq 1 \} \subseteq \mathbb{C}$, and the function \isa{proots\_within p s} gives the set of roots of the polynomial \isa{p} within the set \isa{s}:
\begin{isabelle}
	\isacommand{definition}	\ proots\_within::"'a::comm\_semiring\_0\ poly\ \isasymRightarrow \ 'a\ set\ \isasymRightarrow \ 'a\ set"\ \isakeyword{where}\isanewline
	\ \ "proots\_within\ p\ s\ =\ \isacharbraceleft x\isasymin s.\ poly\ p\ x=0\isacharbraceright "
\end{isabelle}

The next step is to make the definition \isa{no\_proots\_line} executable. This is achieved by proving a \emph{code equation}, where the left-hand side of the equation is the target definition and the right-hand side is an executable expression. In the case of \isa{no\_proots\_line}, the code equation is the following lemma:
\begin{lemma}[\isa{no\_proots\_line\_code[code]}] \label{thm:no_proots_line_code}
	\vspace{-7pt}
	\begin{isabellebody}
		\isanewline
		\ \ "no\_proots\_line\ p\ a\ b\ =\ (if\ poly\ p\ a\ \isasymnoteq \ 0\ \isasymand \ poly\ p\ b\ \isasymnoteq \ 0\ then\ \isanewline
		\ \ \ \ \ \ \ \ \ \ \ \ \ \ \ \ \ \ \ \ \ \ \ \ \ \ \ \ \ \ (let\ p\isactrlsub c\ =\ p\ \isasymcirc \isactrlsub p\ [:a,\ b\ -\ a:];\isanewline
		\ \ \ \ \ \ \ \ \ \ \ \ \ \ \ \ \ \ \ \ \ \ \ \ \ \ \ \ \ \ \ \ \ \ \ p\isactrlsub R\ =\ map\_poly\ Re\ p\isactrlsub c;\isanewline
		\ \ \ \ \ \ \ \ \ \ \ \ \ \ \ \ \ \ \ \ \ \ \ \ \ \ \ \ \ \ \ \ \ \ \ p\isactrlsub I\ =\ map\_poly\ Im\ p\isactrlsub c;\isanewline
		\ \ \ \ \ \ \ \ \ \ \ \ \ \ \ \ \ \ \ \ \ \ \ \ \ \ \ \ \ \ \ \ \ \ \ g\ \ =\ gcd\ p\isactrlsub R\ p\isactrlsub I\isanewline
		\ \ \ \ \ \ \ \ \ \ \ \ \ \ \ \ \ \ \ \ \ \ \ \ \ \ \ \ \ \ \ in\ if\ changes\_itv\_smods\ 0\ 1\ g\ (pderiv\ g)\ =\ 0\ \isanewline
		\ \ \ \ \ \ \ \ \ \ \ \ \ \ \ \ \ \ \ \ \ \ \ \ \ \ \ \ \ \ \ \ \ \ then\ True\ else\ False)\ \isanewline
		\ \ \ \ \ \ \ \ \ \ \ \ \ \ \ \ \ \ \ \ \ \ \ \ \ \ \ else\ False)"
	\end{isabellebody}
\end{lemma}
\noindent where \isa{\isasymcirc \isactrlsub p} is the polynomial composition operation and \isa{map\_poly Re} and \isa{map\_poly Im}, respectively, extract the real and imaginary parts of the complex polynomial \isa{p\isactrlsub c}.
\begin{proof}[Proof of Lemma \ref{thm:no_proots_line_code}]
	Supposing $L : [0,1] \rightarrow \mathbb{C}$ is a linear path from $a$ to $b$:
	$ L(t) = (1-t) a + t b$,
	we know that $p \circ L$ is still a polynomial with complex coefficients. Subsequently, we extract the real and imaginary parts ($p_R$ and $p_I$, respectively) of $p \circ L$ such that
	\[
	p (L (t)) = p_R(t) + i p_I(t).
	\]
	If there is a root of $p$ lying right on $L$, we will be able to obtain some $t_0 \in [0,1]$ such that
	\[
	p_R(t_0) = p_I(t_0) = 0,
	\]
	hence, by letting $g=\gcd(p_R,p_I)$ we have $g(t_0) = 0$. Therefore, the polynomial $p$ has no (complex) root on $L$ if and only if $g$ has no (real) root within the interval $[0,1]$, and the latter can be computationally checked using Sturm's theorem.
\end{proof}

Finally, we define the function \isa{proots\_rectangle} that returns the number of complex roots of a polynomial (counted with multiplicity) within a rectangle defined by its lower left and upper right corner:
\begin{isabelle}
	\isacommand{definition}	\ proots\_rectangle::"complex\ poly\ \isasymRightarrow \ complex\ \isasymRightarrow \ complex\ \isasymRightarrow \ int"\ \isakeyword{where}\isanewline
	\ \ "proots\_rectangle\ p\ a\isactrlsub 1\ a\isactrlsub 3\ =\ proots\_count\ p\ (box\ a\isactrlsub 1\ a\isactrlsub 3)"
\end{isabelle}
where \isa{proots\_count p s} denotes the number of roots of the polynomial $p$ within the set $s$:
\begin{isabelle}
	\isacommand{definition}	\ proots\_count::"'a::idom\ poly\ \isasymRightarrow \ 'a\ set\ \isasymRightarrow \ nat"\ \isakeyword{where}\isanewline
	\ \ "proots\_count\ p\ s\ =\ (\isasymSum r\isasymin proots\_within\ p\ s.\ order\ r\ p)"
\end{isabelle}
The executability of the function \isa{proots\_rectangle} can be established with the following code equation:
\begin{lemma} [\isa{proots\_rectangle\_code1[code]}]
	\vspace{-7pt}
	\begin{isabellebody}
		\isanewline
		\ \ "proots\_rectangle\ p\ a\isactrlsub 1\ a\isactrlsub 3\ =\ \isanewline
		\ \ \ \ (if\ Re\ a\isactrlsub 1\ <\ Re\ a\isactrlsub 3\ \isasymand \ Im\ a\isactrlsub 1\ <\ Im\ a\isactrlsub 3\ then\ \isanewline
		\ \ \ \ \ \ \ \ if\ p\isasymnoteq 0\ then\ \isanewline
		\ \ \ \ \ \ \ \ if\ no\_proots\_line\ p\ a\isactrlsub 1\ (Complex\ (Re\ a\isactrlsub 3)\ (Im\ a\isactrlsub 1))\isanewline
		\ \ \ \ \ \ \ \ \ \ \isasymand \ no\_proots\_line\ p\ (Complex\ (Re\ a\isactrlsub 3)\ (Im\ a\isactrlsub 1))\ a\isactrlsub 3\isanewline
		\ \ \ \ \ \ \ \ \ \ \isasymand \ no\_proots\_line\ p\ a\isactrlsub 3\ (Complex\ (Re\ a\isactrlsub 1)\ (Im\ a\isactrlsub 3))\isanewline
		\ \ \ \ \ \ \ \ \ \ \isasymand \ no\_proots\_line\ p\ (Complex\ (Re\ a\isactrlsub 1)\ (Im\ a\isactrlsub 3))\ a\isactrlsub 1\ then\ \ \isanewline
		\ \ \ \ \ \ \ \ (\isanewline
		\ \ \ \ \ \ \ \ let\ p\isactrlsub 1\ =\ p\ \isasymcirc \isactrlsub p\ [:a\isactrlsub 1,\ Complex\ (Re\ a\isactrlsub 3\ -\ Re\ a\isactrlsub 1)\ 0:];\isanewline
		\ \ \ \ \ \ \ \ \ \ \ \ p\isactrlsub R\isactrlsub 1\ =\ map\_poly\ Re\ p\isactrlsub 1;\ p\isactrlsub I\isactrlsub 1\ =\ map\_poly\ Im\ p\isactrlsub 1;\ g\isactrlsub 1\ =\ gcd\ p\isactrlsub R\isactrlsub 1\ p\isactrlsub I\isactrlsub 1;\isanewline
		\ \ \ \ \ \ \ \ \ \ \ \ p\isactrlsub 2\ =\ p\ \isasymcirc \isactrlsub p\ [:Complex\ (Re\ a\isactrlsub 3)\ (Im\ a\isactrlsub 1),\ Complex\ 0\ (Im\ a\isactrlsub 3\ -\ Im\ a\isactrlsub 1):];\isanewline
		\ \ \ \ \ \ \ \ \ \ \ \ p\isactrlsub R\isactrlsub 2\ =\ map\_poly\ Re\ p\isactrlsub 2;\ p\isactrlsub I\isactrlsub 2\ =\ map\_poly\ Im\ p\isactrlsub 2;\ g\isactrlsub 2\ =\ gcd\ p\isactrlsub R\isactrlsub 2\ p\isactrlsub I\isactrlsub 2;\isanewline
		\ \ \ \ \ \ \ \ \ \ \ \ p\isactrlsub 3\ =\ p\ \isasymcirc \isactrlsub p\ [:a\isactrlsub 3,\ Complex\ (Re\ a\isactrlsub 1\ -\ Re\ a\isactrlsub 3)\ 0:];\isanewline
		\ \ \ \ \ \ \ \ \ \ \ \ p\isactrlsub R\isactrlsub 3\ =\ map\_poly\ Re\ p\isactrlsub 3;\ p\isactrlsub I\isactrlsub 3\ =\ map\_poly\ Im\ p\isactrlsub 3;\ g\isactrlsub 3\ =\ gcd\ p\isactrlsub R\isactrlsub 3\ p\isactrlsub I\isactrlsub 3;\isanewline
		\ \ \ \ \ \ \ \ \ \ \ \ p\isactrlsub 4\ =\ p\ \isasymcirc \isactrlsub p\ [:Complex\ (Re\ a\isactrlsub 1)\ (Im\ a\isactrlsub 3),\ Complex\ 0\ (Im\ a\isactrlsub 1\ -\ Im\ a\isactrlsub 3):];\isanewline
		\ \ \ \ \ \ \ \ \ \ \ \ p\isactrlsub R\isactrlsub 4\ =\ map\_poly\ Re\ p\isactrlsub 4;\ p\isactrlsub I\isactrlsub 4\ =\ map\_poly\ Im\ p\isactrlsub 4;\ g\isactrlsub 4\ =\ gcd\ p\isactrlsub R\isactrlsub 4\ p\isactrlsub I\isactrlsub 4\isanewline
		\ \ \ \ \ \ \ \ in\ \isanewline
		\ \ \ \ \ \ \ \ \ \ -\ (changes\_alt\_itv\_smods\ 0\ 1\ (p\isactrlsub R\isactrlsub 1\ div\ g\isactrlsub 1)\ (p\isactrlsub I\isactrlsub 1\ div\ g\isactrlsub 1)\isanewline
		\ \ \ \ \ \ \ \ \ \ \ \ +\ changes\_alt\_itv\_smods\ 0\ 1\ (p\isactrlsub R\isactrlsub 2\ div\ g\isactrlsub 2)\ (p\isactrlsub I\isactrlsub 2\ div\ g\isactrlsub 2)\isanewline
		\ \ \ \ \ \ \ \ \ \ \ \ +\ changes\_alt\_itv\_smods\ 0\ 1\ (p\isactrlsub R\isactrlsub 3\ div\ g\isactrlsub 3)\ (p\isactrlsub I\isactrlsub 3\ div\ g\isactrlsub 3)\isanewline
		\ \ \ \ \ \ \ \ \ \ \ \ +\ changes\_alt\_itv\_smods\ 0\ 1\ (p\isactrlsub R\isactrlsub 4\ div\ g\isactrlsub 4)\ (p\isactrlsub I\isactrlsub 4\ div\ g\isactrlsub 4))\ div\ 4\isanewline
		\ \ \ \ \ \ \ \ )\isanewline
		\ \ \ \ \ \ \ \ else\ Code.abort\ (STR\ ''proots\_rectangle\ fails\ when\ there\ is\ \isanewline
		\ \ \ \ \ \ \ \ \ \ \ \ \ \ \ \ a\ root\ on\ the\ border.'')\ (\isasymlambda \_.\ proots\_rectangle\ p\ a\isactrlsub 1\ a\isactrlsub 3)\isanewline
		\ \ \ \ \ \ \ \ else\ Code.abort\ (STR\ ''proots\_rectangle\ fails\ when\ p=0.'')\ \isanewline
		\ \ \ \ \ \ \ \ \ \ \ \ \ \ \ \ (\isasymlambda \_.\ proots\_rectangle\ p\ a\isactrlsub 1\ a\isactrlsub 3)\isanewline
		\ \ \ \ \ \ \ \ else\ 0\isanewline
		\ \ \ \ )"
	\end{isabellebody}
\end{lemma}
The proof of the above code equation roughly follows Equations (\ref{eq:roots_rectangle1}) and (\ref{eq:roots_rectangle2}), where \isa{no\_proots\_line} checks if there is a root of \isa{p} on the rectangle's border. Note that the gcd calculations here, such as \isa{g\isactrlsub 1\ =\ gcd\ p\isactrlsub R\isactrlsub 1\ p\isactrlsub I\isactrlsub 1}, are due to the coprime assumption in Lemma 
\ref{thm:cindex_polyE_changes_alt_itv_mods}.

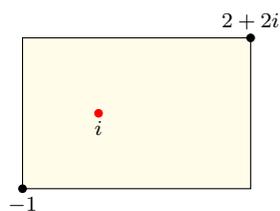
\begin{figure}[ht]
	\centering
	\begin{tikzpicture}[scale=1]

	\fill [yellow!10] (-1,0) rectangle (2,2);
	\draw (-1,0) -- (2,0) -- (2,2) -- (-1,2) -- (-1,0) ;

	\draw [fill,red] (0,1) circle [radius=.5mm];
	\node at (0,1) [below] {$i$};

	\draw [fill] (-1,0) circle [radius=.5mm];
	\node at (-1,0) [below] {$-1$};

	\draw [fill] (2,2) circle [radius=.5mm];
	\node at (2,2) [above] {$2+2 i$};

	\end{tikzpicture}
	\caption{A complex point $i$ and a rectangle defined by its lower left corner $-1$ and upper right corner $2+2 i$}
	\label{fig:roots_rectangle_example}
\end{figure}

\begin{example}
	Given a rectangle defined by $(-1, 2 + 2 i)$ (as illustrated in Fig. \ref{fig:roots_rectangle_example}) and a polynomial $p$ with complex coefficients:
	\[
	p(x) = x^2 - 2 i x - 1 = (x-i)^2
	\]
	we can now type the following command to count the number of roots within the rectangle:
	\begin{isabelle}
		\isacommand{value}		\ "proots\_rectangle\ [:-1,\ -2*\isasymi ,\ 1:]\ (-\isasymi )\ (2+2*\isasymi )"
	\end{isabelle}
	which will return $2$ as $p$ has exactly two complex roots (i.e.
	$i$ with multiplicity 2) in the area.
\end{example}

\subsection{Roots in a Half-plane}

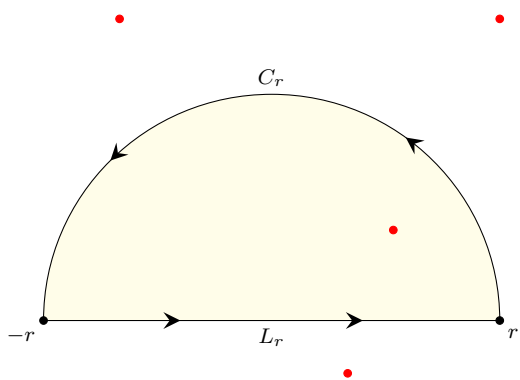
\begin{figure}[ht]
	\centering
	\begin{tikzpicture}[scale=1]
	\fill [yellow!10] (0,0) circle [radius=3];
	\fill [white] (-3,-3) rectangle (3,0);

	\draw (3,0) arc (0:180:3) [arrow inside={end=stealth,opt={,scale=2}}{0.3,0.75}];
	\draw (-3,0) -- (3,0) [arrow inside={end=stealth,opt={,scale=2}}{0.3,0.7}];

	\draw [fill,red] (-2,4) circle [radius=.5mm] node (r1) {};
	\draw [fill,red] (1.6,1.2) circle [radius=.5mm] node (r2) {} ;
	\draw [fill,red] (1,-.7) circle [radius=.5mm] node (r3) {};
	\draw [fill,red] (3,4) circle [radius=.5mm] node (r4) {};

	\draw [fill] (-3,0) circle [radius=.5mm] node [below left] {$-r$};
	\draw [fill] (3,0) circle [radius=.5mm] node [below right] {$r$};

	%

	\node at (0,3) [above] {$C_r$};
	\node at (0,0) [below] {$L_r$};
	\end{tikzpicture}
	\caption{Complex roots of a polynomial (red dots) and a linear path ($L_r$) concatenated by a semi-circular path ($C_r$) on the complex plane}
	\label{fig:roots_halfplane}
\end{figure}

For roots in a half-plane, we can start with a simplified case, where we count the number of roots of a polynomial in the upper half-plane of $\mathbb{C}$:
\begin{isabelle}
	\isacommand{definition}	\ proots\_upper::"complex\ poly\ \isasymRightarrow \ int"\ \isakeyword{where}\isanewline
	\ \ "proots\_upper\ p\ =\ proots\_count\ p\ \isacharbraceleft z.\ Im\ z>0\isacharbraceright "
\end{isabelle}
As usual, our next step is to set up the executability of \isa{proots\_upper}. To achieve that, we first define a linear path $L_r (t) = (1-t) (- r) + t r$ and a semi-circular path $C_r (t) = r e^{i \pi t}$, as illustrated in Fig.~\ref{fig:roots_halfplane}. Subsequently, let
\[
\begin{split}
C_p(r) &=  p \circ C_r \\
L_p(r) &= p \circ L_r,
\end{split}
\]
and by following Equation (\ref{eq:count_roots_intro3}) we have
\begin{equation} \label{eq:roots_halfplane1}
\begin{split}
N_r & = n(p \circ (L_r+C_r),0) \\
& = \Re(n(L_p(r), 0)) + \Re(n(C_p(r), 0))
\end{split}
\end{equation}
where $N_r$ is the number of roots of $p$ inside the path $L_r+C_r$. Note that as $r$ approaches positive infinity, $N_r$ will be the roots on the upper half-plane (i.e., \isa{proots\_upper p}), which is what we are aiming for. For this reason, it is natural for us to examine two cases:
\begin{equation*}
\lim_{r \rightarrow +\infty} \Re(n(L_p(r), 0)) =\ ?
\end{equation*}
\begin{equation*}
\lim_{r \rightarrow +\infty} \Re(n(C_p(r), 0)) =\ ?.
\end{equation*}

For the case of $\lim_{r \rightarrow +\infty} \Re(n(L_p(r), 0))$, we can have
\begin{lemma}[\isa{Re\_winding\_number\_poly\_linepth}]
	\label{thm:Re_winding_number_poly_linepath}
	\vspace{-7pt}
	\begin{isabellebody}
		\isanewline
		\ \ \isakeyword{fixes}\ p::"complex\ poly"\isanewline
		\ \ \isakeyword{defines}\ "L\isactrlsub p\ \isasymequiv \ (\isasymlambda r::real.\ poly\ p\ o\ linepath\ (-r)\ r)"\isanewline
		\ \ \isakeyword{assumes}\ "lead\_coeff\ p=1"\ \isakeyword{and}\ "\isasymforall x\isasymin \isacharbraceleft x.\ poly\ p\ x=0\isacharbraceright .\ Im\ x\isasymnoteq 0"\isanewline
		\ \ \isakeyword{shows}\ "((\isasymlambda r.\ 2*Re\ (winding\_number\ (L\isactrlsub p\ r)\ 0)\ +\ cindex\_pathE\ (L\isactrlsub p\ r)\ 0)\ \isanewline
		\ \ \ \ \ \ \ \ \ \ \ \ \isasymlonglongrightarrow \ 0)\ at\_top"
	\end{isabellebody}
\end{lemma}
\noindent which essentially indicates
\begin{equation}  \label{eq:halfplane_lim1}
\lim_{r \rightarrow +\infty} \Re(n(L_p(r), 0)) = 	- \frac{1}{2} \lim_{r \rightarrow +\infty} \Indp(L_p(r),0),
\end{equation}
provided that the polynomial $p$ is monic and does not have any root on the real axis.

Next, for $\lim_{r \rightarrow +\infty} \Re(n(C_p(r), 0))$, we  first derive a lemma about $C_r$:
\begin{lemma}[\isa{Re\_winding\_number\_tendsto\_part\_circlepath}]
	\label{thm:Re_winding_number_tendsto_part_circlepath}
	\vspace{-7pt}
	\begin{isabellebody}
		\isanewline
		\ \ \isakeyword{fixes}\ z\ z\isactrlsub 0::complex\isanewline
		\ \ \isakeyword{shows}\ "((\isasymlambda r.\ Re\ (winding\_number\ (part\_circlepath\ z\ r\ 0\ pi\ )\ z\isactrlsub 0))\ \isanewline
		\ \ \ \ \ \ \ \ \ \ \ \ \isasymlonglongrightarrow \ 1/2)\ at\_top"
	\end{isabellebody}
\end{lemma}
\noindent that is, $\lim_{r \rightarrow +\infty} \Re(n(C_r, 0)) = 1/2$, following which and by induction we have
\begin{lemma}[\isa{Re\_winding\_number\_poly\_part\_circlepath}]
	\label{thm:Re_winding_number_poly_part_circlepath}
	\vspace{-7pt}
	\begin{isabellebody}
		\isanewline
		\ \ \isakeyword{fixes}\ z::complex\ \isakeyword{and}\ p::"complex\ poly"\isanewline
		\ \ \isakeyword{defines}\ "C\isactrlsub p\ \isasymequiv \ (\isasymlambda r::real.\ poly\ p\ o\ part\_circlepath\ z\ r\ 0\ pi)"\isanewline
		\ \ \isakeyword{assumes}\ "degree\ p>0"\isanewline
		\ \ \isakeyword{shows}\ "((\isasymlambda r.\ Re\ (winding\_number\ (C\isactrlsub p\ r)\ 0))\ \isasymlonglongrightarrow \ degree\ p/2)\ at\_top"
	\end{isabellebody}
\end{lemma}
\noindent which is equivalent to 
\begin{equation} \label{eq:halfplane_lim2}
\lim_{r \rightarrow +\infty} \Re(n(C_p(r), 0)) = \frac{\deg(p)}{2},
\end{equation}
provided $\deg(p) > 0$.

Putting Equations (\ref{eq:halfplane_lim1}) and (\ref{eq:halfplane_lim2}) together yields the core lemma about \isa{proots\_upper} in this section:
\begin{lemma}[\isa{proots\_upper\_cindex\_eq}]
	\label{thm:proots_upper_cindex_eq}
	\vspace{-7pt}
	\begin{isabellebody}
		\isanewline
		\ \ \isakeyword{fixes}\ p::"complex\ poly"\isanewline
		\ \ \isakeyword{assumes}\ "lead\_coeff\ p=1"\ \isakeyword{and}\ "\isasymforall x\isasymin \isacharbraceleft x.\ poly\ p\ x=0\isacharbraceright .\ Im\ x\isasymnoteq 0"\ \isanewline
		\ \ \isakeyword{shows}\ "proots\_upper\ p\ =\isanewline
		\ \ \ \ \ \ \ \ \ \ \ \ \ (degree\ p\ -\ cindex\_poly\_ubd\ (map\_poly\ Im\ p)\ (map\_poly\ Re\ p))/2"
	\end{isabellebody}
\end{lemma}
\noindent where \isa{cindex\_poly\_ubd\ (map\_poly\ Im\ p)\ (map\_poly\ Re\ p)} is mathematically interpreted as $\Ind_{-\infty}^{+\infty} (\lambda t.\, \Im(p(t)) / \Re(p(t)))$, which is derived from $\lim_{r \rightarrow \infty} \Indp(L_p(r),0)$ in Equation (\ref{eq:halfplane_lim1}) since
\[
\begin{split}
\lim_{r \rightarrow +\infty} \Indp(L_p(r),0) & = \lim_{r \rightarrow +\infty} \Indp(L_p(r),0) \\
& = \lim_{r \rightarrow +\infty} \Ind_0^1 \left(\lambda t.\ \frac{\Im(L_p(r,t))}{\Re(L_p(r,t))} \right) \\
& = \lim_{r \rightarrow +\infty} \Ind_{-r}^r \left(\lambda t.\ \frac{\Im(p(t))}{\Re(p(t))} \right) \\
& = \Ind_{-\infty}^{+\infty} \left(\lambda t.\ \frac{\Im(p(t))}{\Re(p(t))} \right). \\
\end{split}
\]

Finally, following Lemma \ref{thm:proots_upper_cindex_eq}, the executability of the function \isa{proots\_upper} is established:
\begin{lemma}[\isa{proots\_upper\_code1[code]}]
	\label{thm:proots_upper_code1}
	\vspace{-7pt}
	\begin{isabellebody}
		\isanewline
		\ \ "proots\_upper\ p\ =\ \isanewline
		\ \ \ \ \ \ (if\ p\ \isasymnoteq \ 0\ then\isanewline
		\ \ \ \ \ \ \ \ (let\ p\isactrlsub m=\ smult\ (inverse\ (lead\_coeff\ p))\ p;\isanewline
		\ \ \ \ \ \ \ \ \ \ \ \ \ p\isactrlsub I=\ map\_poly\ Im\ p\isactrlsub m;\isanewline
		\ \ \ \ \ \ \ \ \ \ \ \ \ p\isactrlsub R=\ map\_poly\ Re\ p\isactrlsub m;\isanewline
		\ \ \ \ \ \ \ \ \ \ \ \ \ g\ =\ gcd\ p\isactrlsub I\ p\isactrlsub R\isanewline
		\ \ \ \ \ \ \ \ \ in\isanewline
		\ \ \ \ \ \ \ \ \ \ \ \ if\ changes\_R\_smods\ g\ (pderiv\ g)\ =\ 0\ \isanewline
		\ \ \ \ \ \ \ \ \ \ \ \ then\ \isanewline
		\ \ \ \ \ \ \ \ \ \ \ \ \ \ (degree\ p\ -\ changes\_R\_smods\ p\isactrlsub R\ p\isactrlsub I)\ div\ 2\ \isanewline
		\ \ \ \ \ \ \ \ \ \ \ \ else\ \isanewline
		\ \ \ \ \ \ \ \ \ \ \ \ \ \ Code.abort\ (STR\ ''proots\_upper\ fails\ when\ there\ is\ a\ root\ \isanewline
		\ \ \ \ \ \ \ \ \ \ \ \ \ \ \ \ on\ the\ border.'')\ (\isasymlambda \_.\ proots\_upper\ p)\ \isanewline
		\ \ \ \ \ \ \ \ )\isanewline
		\ \ \ \ \ \ \ else\ \isanewline
		\ \ \ \ \ \ \ \ \ \ Code.abort\ (STR\ ''proots\_upper\ fails\ when\ p=0.'')\ \isanewline
		\ \ \ \ \ \ \ \ \ \ \ \ (\isasymlambda \_.\ proots\_upper\ p))"
	\end{isabellebody}
\end{lemma}
\noindent where
\begin{itemize}
	\item \isa{smult\ (inverse\ (lead\_coeff\ p))\ p} divides the polynomial \isa{p} by its leading coefficient so that the resulting polynomial \isa{p\isactrlsub m} is monic. This corresponds to the assumption \isa{lead\_coeff\ p=1} in Lemma \ref{thm:proots_upper_cindex_eq}.
	\item \isa{changes\_R\_smods\ g\ (pderiv\ g)\ =\ 0} checks if \isa{p} has no root lying on the real axis, which is due to the second assumption in Lemma \ref{thm:proots_upper_cindex_eq}.
	\item \isa{changes\_R\_smods\ p\isactrlsub R\ p\isactrlsub I} evaluates
	\[
	\Ind_{-\infty}^{+\infty} \left(\lambda t.\, \frac{\Im(p_I(t))}{\Re(p_R(t))} \right)
	\]
	by following Equation (\ref{eq:roots_rectangle2}).
\end{itemize}

As for the general case of a half-plane, we can have a definition as follows:
\begin{isabelle}
	\isacommand{definition}	\ proots\_half::"complex\ poly\ \isasymRightarrow \ complex\ \isasymRightarrow \ complex\ \isasymRightarrow \ int"\ \isakeyword{where}\isanewline
	\ \ "proots\_half\ p\ a\ b\ =\ proots\_count\ p\ \isacharbraceleft w.\ Im\ ((w-a)\ /\ (b-a))\ >\ 0\isacharbraceright "
\end{isabelle}
which encodes the number of roots in the left half-plane of the vector $b-a$. Roots of \isa{p} in this half-plane can be transformed to roots of \isa{p\ \isasymcirc \isactrlsub p\ [:a,\ b-a:]} in the upper half-plane of $\mathbb{C}$:
\begin{lemma}[\isa{proots\_half\_proots\_upper}]
	\label{thm:proots_half_proots_upper}
	\vspace{-7pt}
	\begin{isabellebody}
		\isanewline
		\ \ \isakeyword{fixes}\ a\ b::complex\ \isakeyword{and}\ p::"complex\ poly"\isanewline
		\ \ \isakeyword{assumes}\ "a\isasymnoteq b"\ \isakeyword{and}\ "p\isasymnoteq 0"\isanewline
		\ \ \isakeyword{shows}\ "proots\_half\ p\ a\ b\ =\ proots\_upper\ (p\ \isasymcirc \isactrlsub p\ [:a,\ b-a:])"
	\end{isabellebody}
\end{lemma}
And so we can naturally evaluate \isa{proots\_half} through \isa{proots\_upper}:
\begin{lemma}[\isa{proots\_half\_code1[code]}]
	\label{thm:proots_half_code1}
	\vspace{-7pt}
	\begin{isabellebody}
		\isanewline
		\ \ "proots\_half\ p\ a\ b\ =\ \isanewline
		\ \ \ \ \ \ (if\ a\isasymnoteq b\ then\isanewline
		\ \ \ \ \ \ \ if\ p\isasymnoteq 0\ then\isanewline
		\ \ \ \ \ \ \ \ \ \ proots\_upper\ (p\ \isasymcirc \isactrlsub p\ [:a,\ b\ -\ a:])\ \isanewline
		\ \ \ \ \ \ \ else\ Code.abort\ (STR\ ''proots\_half\ fails\ when\ p=0.'')\ \isanewline
		\ \ \ \ \ \ \ \ \ \ \ \ \ \ (\isasymlambda \_.\ proots\_half\ p\ a\ b)\ \isanewline
		\ \ \ \ \ \ \ else\ 0)"
	\end{isabellebody}
\end{lemma}

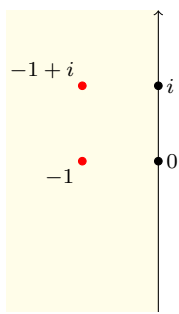
\begin{figure}[ht]
	\centering
	\begin{tikzpicture}[scale=1]

	\fill [yellow!10] (-2,-2) rectangle (0,2);
	\draw [->](0,-2) -- (0,2);

	\draw [fill] (0,0) circle [radius=.5mm] node [right] {$0$};

	\draw [fill] (0,1) circle [radius=.5mm] node [right] {$i$};

	\draw [fill,red] (-1,0) circle [radius=.5mm];
	\node at (-1,0) [below left] {$-1$};

	\draw [fill,red] (-1,1) circle [radius=.5mm];
	\node at (-1,1) [above left] {$-1+i$};

	\end{tikzpicture}
	\caption{Complex roots of a polynomial (red dots) and a vector $(0,i)$}
	\label{fig:roots_halfplane_example}
\end{figure}

\begin{example} \label{ex:roots_left_half}
	We can now use the following command
	\begin{isabelle}
		\isacommand{value}		\ "proots\_half\ [:1-\isasymi ,\ 2-\isasymi ,\ 1:]\ 0\ \isasymi "
	\end{isabelle}
	to decide that the polynomial
	\[
	p(x) = x^2 + (2-i) x + (1-i) = (x+1) (x + 1 -i)
	\]
	has exactly two roots within the left half-plane of the vector $(0,i)$, as shown in Fig. \ref{fig:roots_halfplane_example}.
\end{example}

Despite our naive implementation, both \isa{proofs\_half} and \isa{proots\_rectangle} are applicable for small or medium examples. For most polynomials with coefficient bitsize up to 10 and degree up to 30, our complex root counting procedures terminate within minutes. 

\section{Limitations and Future Work} \label{sec:cindex_limitations}

There are, of course, several improvements that can be made on both the evaluation tactic of \S\ref{sec:tactic_for_evaluation} and the root counting procedures of \S\ref{sec:root_counting}. As the tactic is intended to be applied to winding numbers with variables, full automation with this tactic is unlikely in most cases, but we can always aim for better automation and an enhanced interactive experience for users (e.g., presenting unsolved goals in a more user-friendly way).

Regarding the two root-counting procedures in  \S\ref{sec:root_counting}, a key limitation is that they do not allow cases where any root is on the border. There are two possible solutions to this problem:
\begin{itemize}
	\item To generalise the definition of winding numbers. The current formulation of winding numbers in Isabelle/HOL follows the one in complex analysis:
	\[
	n(\gamma,z) = \frac{1}{2 \pi i} \oint_\gamma \frac{d w}{w - z}
	\]
	which becomes undefined when the point $z$ is on the image of the path $\gamma$. With more general formulations of winding numbers, such as the algebraic version by Eisermann \cite{MichaelEisermann:2012be}, we may be able to derive a stronger version of the argument principle that allows zeros on the border.
	\item To deploy a more sophisticated strategy to count the number of times that the path winds. Recall that the underlying idea in this paper is to reduce the evaluation of winding numbers to \emph{classifications} of how paths cross some line. The Cauchy index merely provides one classification strategy, which we considered simple and elegant enough for formalisation. In contrast, Collins and Krandick \cite{Collins:1992kf} propose a much more sophisticated strategy for such classifications. Their strategy has, in fact, been widely implemented in modern systems, such as Mathematica and SymPy, to count the number of complex roots.
\end{itemize}
Neither of these two solutions are straightforward to incorporate, hence we leave them for future investigation.

Besides rectangles and half-planes, it is also possible to similarly count the number of roots in an open disk and even a sector:
\[
\mathrm{sector}(z_0,\alpha,\beta) = \{z \mid \alpha < \arg(z - z_0) <\beta  \}
\]
where $\arg(\blank)$ returns the argument of a complex number. Informal proofs of root counting in these domains can be found in Rahman and Schmeisser \cite[Chapter 11]{Rahman:2016us}.

\section{Potential Applications} \label{sec:cindex_remarks}

Rahman and Schmeisser's book \cite[Chapter 11]{Rahman:2016us} and Eisermann's paper \cite{MichaelEisermann:2012be} are the two main sources that our development is built upon. Nevertheless, there are still some differences in formulations:
\begin{itemize}
	\item Rahman and Schmeisser formulated the Cauchy index as in Definitions \ref{def:old_jump} and \ref{def:old_ind}, and we used their formulation in our first attempt. However, after we realised the subtleties discussed in \S\ref{sec:subtleties}, we abandoned this formulation and switched to Eisermann's (i.e., Definition \ref{def:ind}). As a result, the root counting procedures presented in this paper are more general than the ones in their book, having fewer preconditions.
	\item Eisermann formulated a winding number $n(\gamma,z_0)$ in a real-algebraical sense where $\gamma$ is required to be a piecewise polynomial path (i.e., each piece from the path needs to be a polynomial). In comparison, $n(\gamma,z_0)$ in Isabelle/HOL follows the classic definition in complex analysis, and places fewer restrictions on the shape of $\gamma$ (i.e., piecewise continuously differentiable is less restrictive than being a piecewise polynomial) but does not permit $z_0$ to be on the image of $\gamma$ (while Eisermann's formulation does). Consequently, Eisermann's root counting procedure works in more restrictive domains (i.e., he only described the rectangle case in his paper) but does not prevent roots on the border.
\end{itemize}
Another point worth mentioning is the difference between informal and formal proofs. In this development, we generally treated their lemma statements as bald facts: we had to discover our own proofs. For instance, when proving Proposition \ref{prop:winding_eq}, we defined an inductive data type for segments and derived an induction rule for it, which was nothing like the informal proof. Such situations also happened when we justified the root counting procedure in a half-plane. Overall, the formal proofs are about 12000 lines.

Interestingly, the root-counting procedure in a half-plane is also related to the stability problems in the theory of dynamical systems. For instance, let $A \in \mathbb{R}^{n \times n}$ be a square matrix with real coefficients and $y : [0,+\infty) \rightarrow \mathbb{R}^n$ be a function that models the system state over time. A linear dynamical system can be described as an ordinary differential equation:
\begin{equation} \label{eq:routh_stable_ex}
\frac{d y(t)}{d t} = A y(t)
\end{equation}
with an initial condition $y(0)=y_0$. The system of (\ref{eq:routh_stable_ex}) is considered stable if all roots of the characteristic polynomial of $A$ lie within the open left half-plane (i.e., $\{z \mid \Re(z) < 0 \}$), and this stability test is usually referred as the Routh-Hurwitz stability criterion \cite[Section 23]{Arnold:1992tf}\cite[Chapter 9]{Marden:1949ui}. As has been demonstrated in Example \ref{ex:roots_left_half}, counting the number of roots in the left half-plane is within the scope of the procedure \isa{proots\_half}. For this reason, we believe that the development in this paper will be beneficial for reasoning about dynamical systems in Isabelle/HOL.

It is worth mentioning that root counting in a rectangle is usually coupled with a classic problem in computer algebra, namely, complex root isolation. The basic idea is to keep bisecting a rectangle (vertically or horizontally) into smaller ones until a sub-rectangle contains exactly one root or none (provided the target polynomial is square-free). Following this idea, it is possible to build a simple and verified procedure for complex root isolation similar to Wilf's algorithm \cite{Wilf:1978fy}: we start with a large rectangle and then repeatedly apply the verified procedure to count roots during the rectangle bisection phase. However, compared to modern complex procedures \cite{Collins:1992kf,Yap:2011in}, this simplistic approach suffers from several drawbacks:
\begin{itemize}
	\item Our root counting procedure is based on remainder sequences, which are generally considered much slower than those built upon Descartes' rule of signs.
	\item Modern isolation procedures are routinely required to deliver isolation boxes whose sizes meet some user-specified limit, hence they usually keep \emph{refining} the isolation boxes even after the roots have been successfully isolated. The bisection strategy still works in the root refinement stage, but dedicated numerical approaches such as Newton's iteration are commonly implemented for efficiency reasons.
	\item Modern isolation procedures sometimes prefer a bit-stream model in which the coefficients of the polynomial are approximated as a bit stream. This approach is particularly beneficial when the coefficients have extremely large bit-width or consist of algebraic numbers.
	\item Modern implementations usually incorporate numerous low-level optimisations, such as hash tables, which are hard to implement as verified procedures in a theorem prover.
\end{itemize}
Therefore, it is unlikely that our verified root counting procedures will ever deliver high performance. Nevertheless, they can be used to certify results from untrusted external root isolation programs, as in the certificate-based approach to solving univariate polynomial problems \cite{Li:2017cg}.

\section{Related Work} \label{sec:related_work}

Formalisations of the winding number (from an analytical perspective) are available in Coq \cite{brunel:LIPIcs:2013:3896}, HOL Light \cite{harrison-complex} and Isabelle/HOL. To the best of our knowledge, our tactic of evaluating winding numbers through Cauchy indices is novel. As both HOL Light and Isabelle/HOL have a relatively comprehensive library of complex analysis (i.e., at least including Cauchy's integral theorem), our evaluation tactic could be useful when deriving analytical proofs in these two proof assistants.

The ability to count the \emph{real} roots of a polynomial only requires Sturm's theorem, so this capability is widely available among major proof assistants including PVS \cite{Narkawicz:2015do}, Coq \cite{Mahboubi:2012gg}, HOL Light \cite{mclaughlin-harrison} and Isabelle \cite{Eberl:2015kb,Sturm_Tarski-AFP,Li:2017cg}. However, as far as we know, our procedures to count \emph{complex} roots are novel, as they require a formalisation of the argument principle \cite{Li_ITP2016}, which is only available in Isabelle at the time of writing.

\section{ Conclusion} \label{sec:conclusion}

In this paper, we have described a novel tactic \isa{winding\_eval}  to evaluate winding numbers via Cauchy indices: given a goal of the form
\[
	n(\gamma_1+\gamma_2+\cdots+\gamma_n,z_0) = k,
\]
the tactic converts the target into an equality about Cauchy indices:
\[
	\Indp(\gamma_1,z_0) + \Indp(\gamma_2,z_0) + \cdots + \Indp(\gamma_n,z_0) = -2 k.
\]
This can be then solved by individually evaluating $\Indp(\gamma_1,z_0)$,...,$\Indp(\gamma_n,z_0)$. As open variables may occur in those Cauchy indices, the evaluation of them is unlikely to be fully automatic, but we provide lemmas (e.g., Lemma 
\ref{thm:cindex_pathE_linepath}) to mitigate the laborious process. The tactic \isa{winding\_eval} has greatly helped us with the motivating proofs shown in~\S\ref{sec:cauchy_motivating_ex}, and we believe that it should be also beneficial in similar situations when dealing with winding numbers in a formal framework.

We have further related Cauchy indices to the argument principle and developed novel verified procedures to count the complex roots of a polynomial within the areas of rectangles and half-planes. Despite the limitations of not allowing roots on the border (which we will solve in future work), the ability to formally count complex roots is believed to lay the foundations for conducting stability analysis (e.g., the Routh-Hurwitz stability criterion) in the framework of the Isabelle theorem prover.

\paragraph*{Acknowledgements.}

We thank Dr. Angeliki Koutsoukou-Argyraki and Anthony Bordg for commenting on the early version of this draft. The work was supported by the ERC Advanced Grant ALEXANDRIA (Project 742178), funded by the European Research Council.

\bibliographystyle{spmpsci}

\bibliography{references}

\end{document}